\documentclass[12pt]{article}
\RequirePackage{amsthm, amsmath, amsfonts, amssymb}
\RequirePackage{mathtools}
\RequirePackage[authoryear]{natbib}
\RequirePackage[pdfencoding=auto,psdextra,colorlinks,citecolor=blue,urlcolor=blue]{hyperref}
\RequirePackage[colorlinks,citecolor=blue,urlcolor=blue]{hyperref}
\RequirePackage{epsfig,longtable,xcolor,algpseudocode}
\RequirePackage{graphicx}
\usepackage{soul}

\usepackage{xr}
\usepackage[linesnumbered, ruled, vlined]{algorithm2e}
\usepackage{multirow}

\usepackage{bbm}

\usepackage{multirow,multicol}
\usepackage{lscape}
\usepackage{float}
\usepackage{makecell}
\usepackage{array}

\usepackage[font={footnotesize,it}]{caption}
\usepackage{subcaption}

\newcommand{\blind}{1}

\newcommand{\bfm}[1]{\ensuremath{\mathbf{#1}}}
\def\ba{\bfm a}     \def\bA{\bfm A}          
\def\bb{\bfm b}               
               
     \def\bD{\bfm D}          
\def\be{\bfm e}          \def\cE{{\cal  E}}     
\def\bff{\bfm f}    \def\bF{\bfm F}     \def\cF{{\cal  F}}     
     \def\bG{\bfm G}          
          \def\cH{{\cal  H}}     
     \def\bI{\bfm I}          
     \def\bJ{\bfm J}          
          \def\cK{{\cal  K}}     
          \def\cL{{\cal  L}}     
               
          \def\cN{{\cal  N}}     
               
\def\bp{\bfm p}     \def\bP{\bfm P}     \def\cP{{\cal  P}}     
\def\bq{\bfm q}          \def\cQ{{\cal  Q}}

          \def\cV{{\cal  V}}

\def\by{\bfm y}     \def\bY{{\bfm Y}}          
\def\bz{\bfm z}     \def\bZ{\bfm Z}

\newcommand{\bfsym}[1]{\ensuremath{\boldsymbol{#1}}}
\def \balpha   {\bfsym{\alpha}}

         \def \btheta   {\bfsym{\theta}}
        
      \def \bmu      {\bfsym{\mu}}
\def \bnu      {\bfsym{\nu}}

\def \bGamma   {\bfsym{\Gamma}}       
\def \bTheta   {\bfsym{\Theta}}       
\def \bXi      {\bfsym{\Xi}}          
\def \bSigma   {\bfsym{\Sigma}}

\DeclareMathOperator{\diag}{diag}

\def \RR	{\mathbb{R}}

\def \PP {\mathbb{P}}

\def \wh{\widehat}
\def \tilde{\widetilde}

\newcommand{\beq}  {\begin{equation}}
\newcommand{\eeq}  {\end{equation}}
\newcommand{\beqn} {\begin{eqnarray}}
\newcommand{\eeqn} {\end{eqnarray}}
\newcommand{\beqnn}{\begin{eqnarray*}}
\newcommand{\eeqnn}{\end{eqnarray*}}

\numberwithin{equation}{section}
\theoremstyle{plain}
\newtheorem{thm}{Theorem}[section]

\newtheorem{cor}{Corollary}[section]

\newtheorem{prop}{Proposition}[section]
\newtheorem{cond}{Condition}[section]

\newtheorem{rem}{Remark}[section]

\begin{document}

\def\spacingset#1{\renewcommand{\baselinestretch}%
{#1}\small\normalsize} \spacingset{1}


\if1\blind
{
  \title{\bf A General Latent Embedding Approach for Modeling Non-uniform High-dimensional Sparse Hypergraphs with Multiplicity}
  \author{Shihao Wu, Gongjun Xu, and Ji Zhu \\
    Department of Statistics, University of Michigan
    }
    \date{}
  \maketitle
} \fi

\if0\blind
{
  \bigskip
  \bigskip
  \bigskip
  \begin{center}
    {\LARGE\bf A General Latent Embedding Approach for Modeling Non-uniform High-dimensional Sparse Hypergraphs with Multiplicity}
\end{center}
  \medskip
} \fi

\bigskip
\begin{abstract}
Recent research has shown growing interest in modeling hypergraphs, which capture polyadic interactions among entities beyond traditional dyadic relations. 
   However, most existing methodologies for hypergraphs face significant limitations, including their heavy reliance on uniformity restrictions for hyperlink orders and their inability to account for repeated observations of identical hyperlinks. In this work, we introduce a novel and general latent embedding approach that addresses these challenges through the integration of latent embeddings, vertex degree heterogeneity parameters, and an order-adjusting parameter. Theoretically, we investigate the identifiability conditions for the latent embeddings and associated parameters, and we establish the convergence rates of their estimators along with asymptotic distributions.  Computationally, we employ 
   a projected gradient ascent algorithm for parameter estimation. Comprehensive simulation studies demonstrate the effectiveness of the algorithm and validate the theoretical findings. Moreover, an application to a co-citation hypergraph illustrates the advantages of the proposed method.
\end{abstract}

\noindent%
\vfill

\newpage
\spacingset{1.8} 

\section{Introduction}
Network analysis explores relations among entities. Most existing network analysis tools focus on analyzing dyadic relations, i.e., those between two entities, through the examination of graphs 
\citep{newman2018networks}. However, real-life relations often go beyond pairwise interactions. 
For instance, in co-authorship data \citep{ji2016coauthorship,ji2022co}, researchers typically form groups larger than pairs when co-authoring papers. 
In co-sponsorship data \citep{fowler2006connecting}, legislators endorsing legislation as co-sponsors form clusters that usually involve more than two people. 
In gene interaction data \citep{razick2008irefindex}, subsets of genes, rather than just pairs, associate to potentially form protein complexes. Modeling and analyzing these higher-order relations, as highlighted by \cite{ke2019community},  yield deeper insights that are often overlooked by traditional graph analysis.
These higher-order interactions can be naturally captured and aggregated by hypergraphs, which allows for a more comprehensive analysis.

We denote a hypergraph by $\cH(\cV_n,\cE_m)$, where $\cV_n = \{ v_1,v_2,\cdots, v_n\}$ represents the collection of $n$ vertices (individuals) and {$\cE_m = \{e_1,e_2,\cdots,e_m\}$} represents the collection of  $m$ hyperlinks/hyper-edges (interactions). For notation simplicity, we denote $[n] =\{1,\cdots,n\}$ and $\cV_n = [n]$. Each hyperlink $e_j$, for $j\in[m]$, consists of a subset of vertices that form the $j$-th hyperlink; that is, $e_j\subseteq[n]$ with $i\in e_j$ if and only if vertex $i$ is part of the $j$-th hyperlink. For example, if vertices $1,2$ and $3$ form the first hyperlink, we have $e_1 = \{1,2,3\}$. A hyperlink consisting of $d$ vertices is termed an order-$d$ hyperlink. If all hyperlinks $e_j\in\cE_m$ have the same order $d$, the hypergraph $\cH(\cV_n,\cE_m )$ is termed a $d$-uniform hypergraph. For instance, a typical graph network is essentially a 2-uniform hypergraph without repetitive hyperlinks. Conversely, if the hyperlinks in $\cE_m$ vary in their orders, the hypergraph is described as non-uniform. Non-uniform hypergraphs are ubiquitous in the real world. For example, consider co-authorship relationships, where authors are vertices and each paper is treated as a hyperlink that includes its co-authors. The number of co-authors for each paper, i.e., the order of the hyperlink, naturally varies across different papers. 
Furthermore, it is common for the same set of co-authors to collaborate on multiple papers, leading to identical hyperlinks appearing multiple times within a hypergraph. This phenomenon is referred to as the multiplicity of hyperlinks. The non-uniformity and multiplicity of hyperlinks pose unique challenges in modeling hypergraphs.


In this work,  we introduce a latent embedding framework to model general hypergraphs that accommodates the non-uniformity and multiplicity of hyperlinks
under a high-dimensional setup where both $n$ and $m$ go to infinity. 
Assume that each vertex is associated with a latent embedding vector (or vertex embedding), denoted by $\{\bz_i\}_{i\in[n]}\subset \RR^{K}$, where $K$ represents the dimension of the latent embedding space (vertex space). Additionally, each hyperlink also carries a corresponding latent embedding (or hyperlink embedding), denoted by $\{\bff_j\}_{j\in[m]}\subset\RR^{K}$, potentially residing in another latent space (the hyperlink space). Conditioned on $\{\bz_i\}_{i\in[n]}$ and $\{\bff_j\}_{j\in[m]}$, our model assumes that the appearances of each node on the hyperlinks are independent.  In other words, interactions among vertices are captured by these latent embeddings. Such conditional independence, given (hyper)graph embeddings, has been considered in many (hyper)graph models, including the stochastic block model \citep{holland1983stochastic}, the latent space network model \citep{hoff2002latent}, and the  hypergraph degree-corrected block model \citep{ke2019community}. 
Specifically, consider a \emph{hypothetical} hypergraph $\cH([n],\{E_1,\ldots, E_m\})$, where each $E_j$ is a random subset of $[n]$. The \emph{observed} hypergraph $\cH([n],\{e_1,\ldots, e_m\})$ is a realization of $\cH([n],\{E_1,\ldots, E_m\})$. Given the latent embeddings, we assume that $\{E_1,\ldots,E_m\}$ are independent with the following distribution: 
\beq\label{eq:hyper_model}
   \PP(E_j = {e}|\bXi_{m,n}) = \prod_{i\in {e}}p_{ji}\prod_{i\in[n]\setminus {e}} (1-p_{ji}), ~~~
   \text{for any }{e}\subseteq[n],
\eeq
where 
$p_{ji} =  g\big(\beta_{m,n} + \alpha_i + \cK( \bff_j, \bz_i  ) \big)\in$~$(0,1)$ and $\bXi_{m,n} = \big\{ \beta_{m,n},\{\alpha_i\}_{i=1}^n,\{\bz_i\}_{i=1}^n,\{\bff_j\}_{j=1}^m  \big\}$. Here, $\cK(\cdot,\cdot): \RR^{K}\times\RR^{K}\to \RR$ measures the ``interaction'' (to be defined) between the two latent embeddings, and $g(\cdot):\RR \to [0,1]$, a monotonically increasing function, maps
a real value to a probability. 
The $\{\alpha_i\}_{i\in[n]}$ parameters (with $\alpha_i \in \RR$) account for heterogeneity in vertex degrees, which can be interpreted as a popularity measure of the vertices, and
$\beta_{m,n}\in\RR$, the order-adjusting parameter, controls the overall order of hyperlinks. Unlike most existing hypergraph modeling approaches where the order of hyperlinks is uniform, our model accommodates varying hyperlink orders.  However, the orders of hyperlinks typically remain much smaller than both the number of vertices within the hypergraph and the number of observed hyperlinks, which poses an additional modeling challenge. $\beta_{m,n}$ serves to address this challenge by monitoring the overall order of hyperlinks compared to $n$ and $m$. 
In model \eqref{eq:hyper_model}, 
$\{\bz_i\}_{i\in[n]}$ and $\{\bff_j\}_{j\in[m]}$ can be  treated either as realizations of some embedding distribution or as parameters to estimate. We treat them as fixed parameters in this work. In situations where these embeddings are treated as random vectors, our analysis can be regarded as making statistical inferences conditional on them. Moreover, it is worth noting that different $E_j$'s have different distributions. This means that rather than assuming an identical generating mechanism for all $m$ hyperlinks, our framework permits individualized probability assignments for each observed hyperlink, which is more flexible in modeling real-world hypergraphs. More discussion and interpretation of \eqref{eq:hyper_model}  are in Section \ref{sec:model}. 

\subsection{Related work}



Modeling {general} hypergraphs has been recognized as both an important task and a significant challenge. A commonly employed strategy is to project the hypergraph onto a weighted graph, enabling analysis using methods such as modularity maximization \citep{kumar2018hypergraph}, graph-likelihood-based approaches \citep{lee2020robust}, and spectrum-based techniques \citep{ghoshdastidar2015provable} on the converted graph.  These approaches have been applied in the studies of academic collaborations \citep{newman2001structure,ji2016coauthorship} and congress voting relations \citep{fowler2006connecting,lee2017time}. For instance, in a co-authorship hypergraph, a hyperlink consists of all co-authors of a paper, usually more than two. Transforming the hypergraph into a weighted graph involves considering each pair of authors and constructing pairwise links with weights proportional to the number of papers they co-authored. However, as \cite{ke2019community} pointed out, such projection process results in information loss and may yield sub-optimal performance when applied to community detection tasks.

Recent research has seen a surge in the direct modeling and analysis of hypergraphs. {For instance, \cite{stasi2014beta} introduced the hypergraph $\beta$-model to capture node-degree heterogeneity in hypergraph networks. \cite{nandy2024degree} subsequently conducted a rigorous theoretical study of the hypergraph $\beta$-model, establishing convergence rates for its maximum likelihood estimator and detection thresholds for likelihood ratio tests with their optimality demonstrated \citep{chatterjee2011random,mukherjee2018detection}.}
\cite{ke2019community} proposed a regularized tensor power iteration algorithm, called Tensor-SCORE, for community detection on uniform hypergraphs.
Extensions to non-uniform hypergraphs were suggested without further investigation. The hypergraph degree-corrected block model introduced by \cite{ke2019community} incorporates vertex degree heterogeneity into the framework of the hypergraph planted partition model \citep{ghoshdastidar2017consistency}. 
\cite{yuan2021high} developed a tensor-based joint network embedding method that encodes both pairwise links and a subgroup of uniform hyperlinks into a latent space, capturing the dependency between pairwise and multi-way links in predicting potential unobserved hyperlinks.
\cite{yuan2022testing} conducted an extensive study on hypergraph testing for uniform hypergraphs under a spectrum of hyperlink probability scenarios, providing comprehensive theoretical analysis and significant insights into the subject. 
\cite{zhen2022community} introduced a null vertex to augment a non-uniform hypergraph into a uniform hypergraph for community detection purposes.
However, most existing hypergraph modeling approaches have significant limitations, including a heavy reliance on uniform hypergraph structures and the inability to account for repeated observations of identical hyperlinks due to the use of adjacency tensor representation. In the co-authorship data example,  this means that only combinations of authors of a fixed size can be effectively modeled, and information about the number of papers a group of authors has co-authored is lost. In reality, the number of co-authors can vary for different papers, and the same set of co-authors can write more than one paper. 
Furthermore, the handling of non-uniform hypergraphs remains ad hoc. {A recent work \cite{turnbull2024latent} introduced a Bayesian latent space hypergraph model with a posterior sampling scheme  for non-uniform hypergraphs, which does not scale to the high-dimensional hypergraphs considered in our setting and cannot handle hyperlink multiplicity.}
In our view, there is a pressing demand for a general modeling framework {for high-dimensional non-uniform hypergraphs with multiplicity}.

\subsection{Contributions}
Our contribution in this article can be summarized in two main aspects. 
First, we introduce a novel and general approach that addresses the limitations of most existing methods in hypergraph modeling. 
Our approach does not require the stringent uniformity restriction on hyperlinks that is often presumed in existing methods, and it accounts for repeated observations of identical hyperlinks, which are overlooked in many current approaches.
These advantages distinctly set our methodology apart from existing hypergraph modeling approaches. 
 
Second, we investigate the identifiability conditions for the embeddings and associated parameters, and establish convergence rates for their estimators along with their asymptotic distributions. 
In Section \ref{sec:cmle}, we introduce estimators that achieve $F$-consistency for the ``${\bTheta}$'' parameter (to be defined),
as well as for the vertex popularity parameters and the order-adjusting parameter.
In Section \ref{sec:pmle}, we address the challenge of estimating individual components in $\bXi_{m,n}$.  Additional identifiability conditions are necessary to disentangle individual parameters, and we design a penalty term based on these conditions. Notably, this penalty term is not for shrinkage or regularization purposes, but for the implementation of a Lagrangian method to address identifiability issues. Furthermore, we allow the order-adjusting parameter $\beta_{m,n}$ to diverge to $-\infty$ to accommodate the sparsity of the order of hyperlinks. 
This divergence results in irregular behavior of the likelihood function. We dedicate substantial effort to understanding this effect. Given that $\beta_{m,n}$ is designed to control the rate of hyperlink orders in the model, which will be further illustrated in Section \ref{sec:model}, this exploration offers valuable insights into the types of datasets that are well-suited to our methodology. 
As a result, we establish entry-wise consistent estimators for individual parameters in $\bXi_{m,n}$ and characterize their asymptotic distributions. 
These theoretical results subsequently permit estimation and inference on the probability terms $p_{ji}$'s.

\subsection{Organization of the paper and notations}
The remainder of this article is organized as follows. Section \ref{sec:model} provides a detailed introduction and interpretation of the order-adjusted latent embedding approach. Section \ref{sec:stats} investigates the identifiability conditions and estimation methods for the parameters and presents the corresponding statistical theory. 
Section \ref{sec:numerical} presents numerical results on synthetic hypergraphs. Section \ref{sec:author} applies our proposed methodology to a co-citation hypergraph, which leads to interesting discoveries. Section \ref{sec:discussion} concludes the paper with a discussion. The supplementary material includes proofs of the main results, technical lemmas and their proofs, and additional numerical results.

Throughout this work, we use regular letters, bold regular letters, and bold capital letters to denote scalars,  vectors, and matrices, respectively. For any $a,b\in\RR$, we denote their maximum and minimum by $a\vee b = \max(a,b)$ and $a\wedge b = \min(a,b)$, respectively. Given any two sequences $(a_{m,n}), (b_{m,n})$ valued in $\RR$, we say $a_{m,n} \lesssim b_{m,n}$ (or $a_{m,n} = O(b_{m,n})$,  $b_{m,n}\gtrsim a_{m,n}$) if there exists a universal constant $C > 0$ such that $|a_{m,n}| \le C|b_{m,n}|$. We say $a_{m,n}\asymp b_{m,n} $ if $a_{m,n}\lesssim b_{m,n}$ and $a_{m,n}\gtrsim b_{m,n}$. We write $a_{m,n}\ll b_{m,n}$ (or $a_{m,n}=o(b_{m,n})$) if $\lim_{m,n\to\infty} a_{m,n}/b_{m,n} = 0 $ and  $a_{m,n}\gg b_{m,n}$ (or $a_{m,n}=\Omega(b_{m,n})$) if $\lim_{m,n\to\infty} a_{m,n}/b_{m,n} = \infty $. We use $O_p(\cdot)$, $\Omega_p(\cdot)$, and $o_p(\cdot)$ as the stochastic versions of $O(\cdot)$, $\Omega(\cdot)$, and $o(\cdot)$ for convergence in probability.  For any positive integer $a$, we use $[a]$ to denote the set $\{1, 2, \ldots, a\}$. For any column vector $\ba$ and matrix $\bA$, we use $\ba^{\top}$ and $\bA ^ {\top}$ to denote their transposes, respectively. For an event $\mathfrak{E}$, the indicator function $\mathbf{1}_{\mathfrak{E}}=1$ if $\mathfrak{E}$ occurs and $0$ otherwise.  For a vector $\ba=(a_1,\cdots,a_n)^{\top}$, we use $\|\ba\|_2$ to denote the standard Euclidean norm, $\|\ba\|_0=\sum_{i\in[n]}\mathbf{1}_{\{a_i\ne 0\}}$ to denote the $L_0$ norm (the number of non-zeros), and $\|\ba\|_{\infty} = \max_i|a_i|$ to denote the infinity/maximum/supreme norm. For a matrix $\bA=(a_{ji})_{m\times n}$, we use $\|\bA\|_F$ and $\|\bA\|_2$ to denote its Frobenius and spectral norms, respectively. We use $\|\bA\|_{\max} = \max_{j\in[m],i\in[n]}|a_{ji}|$ to denote the matrix maximum/supreme norm. For a square matrix $\bA$, we use $\rho_{\max}(\bA)$ and $\rho_{\min}(\bA)$ to denote its maximal and minimal eigenvalues, respectively.

\section{The order adjusted latent embedding approach}\label{sec:model}


In this paper, we consider $\cK(\cdot,\cdot)$ in \eqref{eq:hyper_model} as the Euclidean inner product in a $K$-dimensional space and employ the logistic function as $g$. 
Specifically, for $\ba,\bb\in\RR^K$, $\cK(\ba,\bb) =\ba^{\top}\bb = \sum_{k=1}^Ka_kb_k$, and for $a\in\RR$, $g(a) = \sigma(a):= 1/\big(1+\exp(-a)\big)$. 
Then we have
\beq\label{eq:hyper0}
  p_{ji} = \sigma(\beta_{m,n} + \alpha_i + \bff_j^\top\bz_i).  
\eeq
We show in this section that using the Euclidean inner product in our model offers a clear and coherent interpretation. Although we focus on the logistic link function in this article, it is worth noting that our theoretical analysis is adaptable and can be effectively applied to a variety of other link functions that exhibit similar characteristics.  In contrast to conventional approaches that often focus on modeling the latent embeddings of vertices, with hyperlinks being generated from these vertex embeddings, our method jointly models latent embeddings of vertices and hyperlinks.  This joint modeling framework offers more flexibility and a more comprehensive understanding of the hyperlink embeddings.

One concern with the model \eqref{eq:hyper0} is the identifiability of the parameters, primarily because the probability value in \eqref{eq:hyper_model} is determined by $p_{ji}$, which remains invariant under certain transformations involving $\beta_{m,n},\alpha_i,\bff_j$ and $\bz_i$. To identify $\beta_{m,n}$ and $\alpha_i$'s,  we impose the constraints $\sum_{i=1}^n \alpha_i = 0$ and $\sum_{j=1}^m \bff_j = \mathbf{0}$. 
The rationale for these restrictions is derived from the right-hand side of \eqref{eq:hyper0}.
  If we add a constant to all $\alpha_i$'s and simultaneously subtract it from $\beta_{m,n}$, the resulting probability distribution remains unchanged. Similarly, adding an arbitrary vector $\bmu_\bff\in\RR^K$ to each $\bff_{j}$ while adjusting each $\alpha_i$ by subtracting $\bmu_{\bff}^{\top}\bz_i$ also preserves the original probability distribution.  We defer the discussion on  the identification of $\bff_j$'s and $\bz_i$'s to Section \ref{sec:pmle}, where we aim to estimate individual parameters in $\bXi_{m,n}$.
Throughout this article, we assume that 
\beq\label{eq:constr_finite}
 \max\big\{\max_{i\in[n]}|\alpha_i|, \max_{j\in[m]}\|\bff_{j}\|_2, \max_{i\in[n]}\|\bz_i\|_2\big\}\le C
 \eeq
 for some constant $C>0$. We allow $\beta_{m,n}$ to diverge to $-\infty$ as $(m,n)\to\infty$, thus $p_{ji}$'s can be arbitrarily close to $0$ in the asymptotic regime. Consequently, the expected order of hyperlinks (i.e. the expected number of vertices in a hyperlink) can be sublinear with respect to the size of the hypergraph.  This is consistent with the observation in practice that the order of hyperlinks  often increases at a much slower rate compared to the growth of the number of vertices.  


To further understand \eqref{eq:hyper0}, consider a specific hyperlink indexed by $j\in[m]$.
For a subset of vertices $e \subset[n]$ with $|{e}| = d$, given $\bXi_{m,n}$, model \eqref{eq:hyper_model} along with \eqref{eq:hyper0} specifies the conditional probability that these vertices constitute hyperlink $j$ as 
\beq\label{eq:probability}
\begin{aligned}
    & \PP\big(E_j= e|\bXi_{m,n}   \big) \\ = &  \prod_{ i\in e } \sigma( \beta_{m,n} + \alpha_i + \bff_j^{\top}\bz_i ) \prod_{i\in[n]\setminus e}\big\{ 1 -   \sigma( \beta_{m,n} + \alpha_i + \bff_j^{\top}\bz_i ) \big\} 
    \\  = & \prod_{i\in {e}   } \frac{\exp(\beta_{m,n} + \alpha_i + \bff_j^{\top}\bz_i)}{1+\exp(\beta_{m,n} + \alpha_i + \bff_j^{\top}\bz_i)}  \cdot  \prod_{i\in [n]\setminus {e} }  \frac{ 1 }{1+\exp(\beta_{m,n} + \alpha_i + \bff_j^{\top}\bz_i)}  \\
    = & \frac{\prod_{i\in {e}}  \exp(\beta_{m,n} + \alpha_i + \bff_j^{\top}\bz_i)     }{ \prod_{i\in[n]} \big\{  1+\exp(\beta_{m,n} + \alpha_i + \bff_j^{\top}\bz_i)   \big\}    } \\
    = & \frac{ \big[\exp\big\{\beta_{m,n} + \sum_{i\in  {e}}  \alpha_i /d + \bff_j^{\top} (\sum_{i\in  {e}}  \bz_i/d)\big\} \big]^d   }{\prod_{i\in[n]} \big\{  1+\exp(\beta_{m,n} + \alpha_i + \bff_j^{\top}\bz_i)   \big\}     }.
\end{aligned}
\eeq
The denominator of this probability remains constant regardless of the specific subset of  vertices ${e}$.  
The numerator depends on a combination of factors: the order-adjusting parameter $\beta_{m,n}$,  the vertices' average popularity $\sum_{i\in {e}}  \alpha_i /d $, the inner product between the hyperlink's embedding $\bff_j$ and the centroid of the vertices' embeddings $\sum_{i\in {e}}  \bz_i/d$,  and the order of the hyperlink $d$. 
The multiplicative operations in the second line of \eqref{eq:probability} suggest a form of \emph{conditional independence} among the vertices  given the embeddings.
Such conditional independence has been widely considered in (hyper)graph network models, for example, by
\cite{holland1983stochastic}, \cite{hoff2002latent}, \cite{ke2019community}, and \cite{ma2020universal}. In this work, we do not model the distributions of embeddings as they can be general. Below, 
\begin{figure}
      \centering
      \setlength\tabcolsep{1pt}
		\renewcommand{\arraystretch}{1}      
      \begin{tabular}{cc}
      \includegraphics[height=5.15cm]{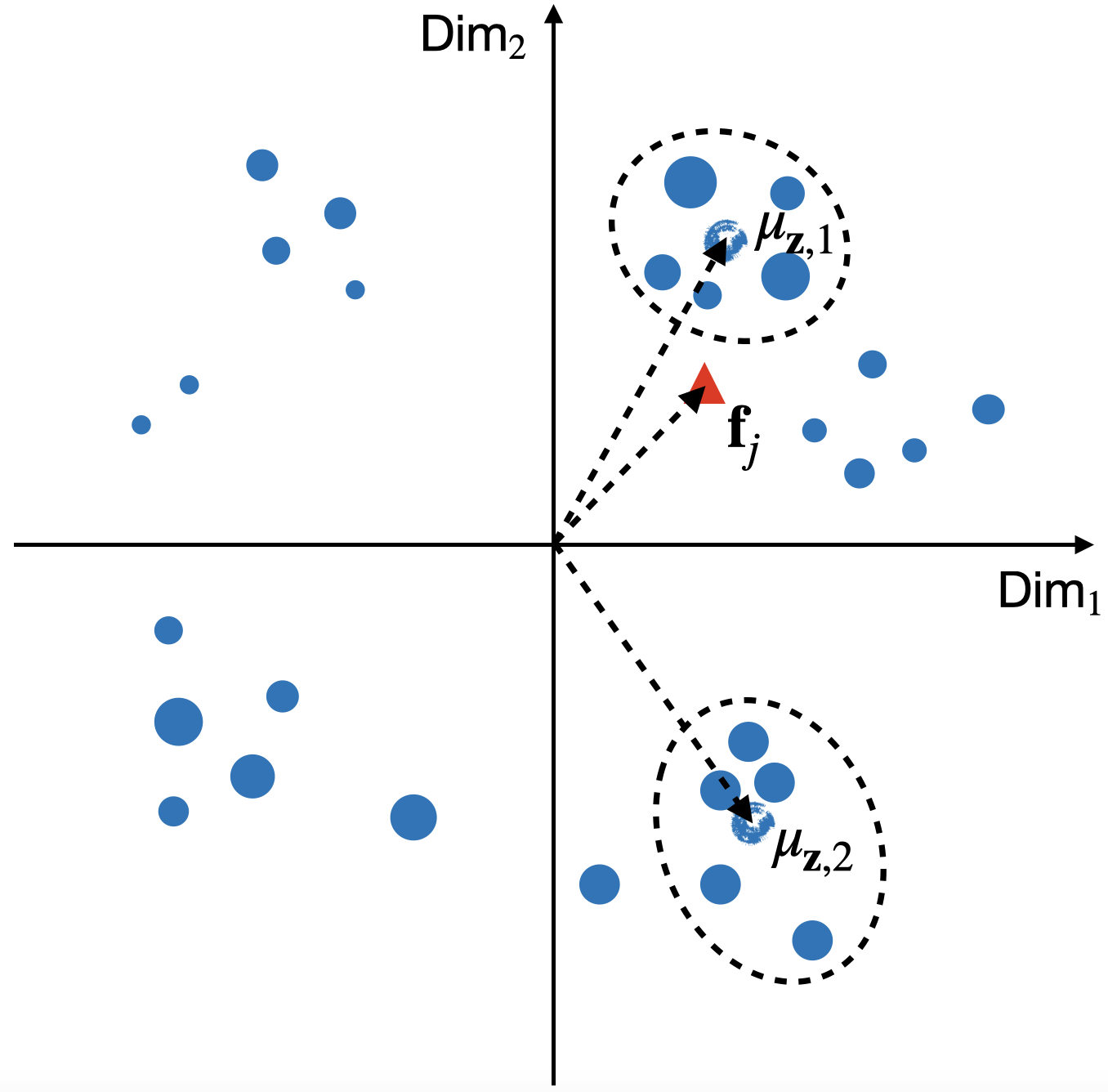}  & \includegraphics[height=5.17cm]{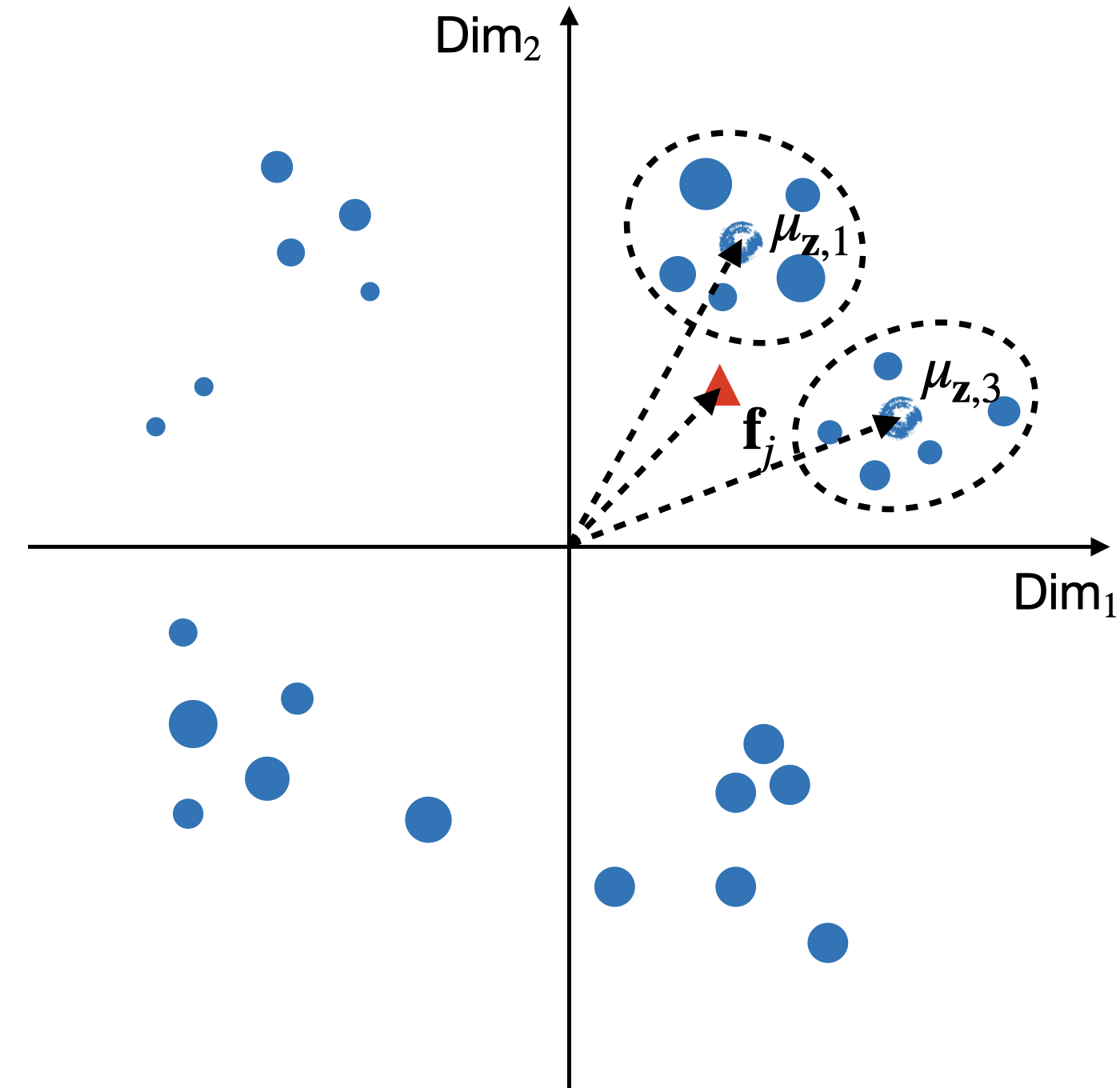}  \\
       (a) & (b)
      \end{tabular}
      \caption{An example of the latent embedding vectors when $K=2$. Figures (a) and (b) display the same latent embedding vectors but consider different  combinations of vertices on hyperlinks. The sizes of the dots correspond to different vertex degree parameters: the larger the dot, the higher its corresponding $\alpha$ value.} 
      \label{fig:embedding0}
\end{figure}
with the hyperlink embedding vector $\bff_j$ being fixed, we use 
Figure \ref{fig:embedding0} to illustrate how the vertex embeddings and associated parameters affect the probability term \eqref{eq:probability}:
\begin{itemize}
\item Positioning of vertices' centroid: Figure \ref{fig:embedding0} (a) highlights two groups of vertices, each with their respective centers denoted as $\bmu_{\bz,1}$ and $\bmu_{\bz,2}$. The Euclidean inner product between $\bmu_{\bz,1}$ and $\bff_j$ is larger than that between $\bmu_{\bz,2}$ and $\bff_{j}$. Focusing only on the term $\bff^{\top}_j(\sum_{i\in{e}} \bz_i/ d)$,  the group of vertices centered around $\bmu_{\bz,1}$ tends to have a higher probability of forming hyperlink $j$ than the other group.
\item Average popularity of vertices: In Figure \ref{fig:embedding0} (b), $\bff_j^{\top}  \bmu_{\bz,1}$ and $\bff_j^{\top}\bmu_{\bz,3} $ are close, and the two groups have the same number of vertices. We consider the term $\sum_{i\in  {e}}  \alpha_i/d $, which we refer to as the average popularity of vertices in ${e}$. Note that larger dot sizes correspond to higher $\alpha$ values in Figure \ref{fig:embedding0}. Due to their higher average popularity, the group of vertices centered around $\bmu_{\bz,1}$ has a higher probability of forming hyperlink $j$ than the group centered around $\bmu_{\bz,3}$.   
\end{itemize}

The order-adjusting parameter $\beta_{m,n}$ is not visualized in Figure \ref{fig:embedding0}, but it is particularly crucial for determining the orders of hyperlinks. 
In the considered sparse scenario, where $\beta_{m,n}\to-\infty$ as $m,n\to\infty$, it holds that  $p_{ji} \asymp \exp(\beta_{m,n})$. This paper studies in details how the divergence rate of $\beta_{m,n}$ affects the hypergraph structure as well as the corresponding estimation and inference procedures.

{
\begin{rem}
Our model uses hyperlink-vertex interactions to model hypergraph networks, whereas existing latent space models typically focus on vertex-vertex interactions. This hyperlink-vertex interaction modeling strategy enables us to move beyond the traditional framework and address the challenges posed by non-uniformity and multiplicity in general hypergraphs, while still preserving several key properties of existing models using vertex-vertex interactions.
We refer the reader to Section~1 of the supplementary material for a detailed discussion. 
\end{rem}
}

\subsection{An initial study of $\beta_{m,n}$ }\label{sec:first_gl}
In this section, we present a simple analysis of how the rate of $p_{ji}$ changes the hypergraph structure as an initial study of $\beta_{m,n}$.  Let $\mathrm{Deg}(i):= \sum_{j\in[m]} \mathbf{1}_{\{i\in e_j\}}$ denote the number of hyperlinks that contain vertex $i$.  We consider the following two concepts:
\begin{itemize} 
\item \textbf{Null vertex:} We call node $i$ a null vertex if $\mathrm{Deg}(i)=0$: if a vertex does not appear in any hyperlink in the observed hypergraph, we cannot learn anything about it.
\item \textbf{Informative link:}  We call hyperlink $j$ an informative link if $|e_j|\ne 0$.  When $|e_j|=0$, hyperlink $j$ does not provide any information for modeling the hypergraph nodes.
\end{itemize}

The following proposition shows that when $p_{ji}$'s are too small, an observed hypergraph generated from \eqref{eq:hyper_model} contains null vertices and non-informative hyperlinks with high chance. 
\begin{prop}[Necessary rate of $p_{ji}$]\label{prop:nece_spar}
 In model \eqref{eq:hyper_model}, suppose $p_{ji} \le \frac{a}{n}$ for all $j\in[m],i\in[n]$, and some constant $a>0$. Then 
  $
  \lim_{m\to\infty}\lim_{n\to\infty}\PP\big(\exists j\in[m], \text{ s.t. } |e_j| = 0   \mid \bXi_{m,n}  \big) = 1$.
  By symmetry, suppose $p_{ji} \le \frac{a}{m}$  for all $j\in[m],i\in[n]$, and some constant $a>0$ in \eqref{eq:hyper_model}. Then 
  $
  \lim_{n\to\infty}\lim_{m\to\infty}\PP\big(\exists i\in[n], \text{ s.t. }\mathrm{Deg}(i) = 0 \mid \bXi_{m,n} \big) = 1$.
\end{prop}

The next proposition shows that the above rate of $p_{ji}$ is almost sufficient (up to a logarithmic factor) to ensure that there is no null vertex and non-informative hyperlink. 

\begin{prop}[Sufficient rate of $p_{ji}$]\label{prop:suff_spar}
In model \eqref{eq:hyper_model}, suppose  $p_{ji} \ge \frac{n^{\epsilon}}{n}$  for all $j\in[m],i\in[n]$ with $n^{\epsilon} = \Omega(\log m)$ for some $\epsilon\in(0,1)$ as $m,n\to \infty$. Then
$
\lim_{m,n\to\infty}\PP\big(|e_{j}| \ne 0,~ \forall j\in[m] \mid \bXi_{m,n} \big) = 1$.
By symmetry, suppose $p_{ji} \ge \frac{m^{\epsilon}}{m}$  for all $j\in[m],i\in[n]$ with $m^{\epsilon} = \Omega(\log n)$ for some $\epsilon\in(0,1)$ as $m,n\to \infty$. Then 
$
\lim_{m,n\to\infty}\PP\big( \mathrm{Deg}(i) \ne 0,~ \forall i\in[n]  \mid \bXi_{m,n}\big) = 1$.
\end{prop}

Recall that $p_{ji}  \asymp \exp(\beta_{m,n})$ when $\beta_{m,n}\to-\infty$. Propositions \ref{prop:nece_spar} and \ref{prop:suff_spar} indicate the required rate of $\beta_{m,n}$ to ensure that a hypergraph avoids null vertices and non-informative links. The established rate is $\exp(\beta_{m,n})\asymp p_{ji}\gtrsim \epsilon_{m,n}/(m\wedge n)$ for some slowly diverging $\epsilon_{m,n}>0$ as $m,n\to\infty$. In Section \ref{sec:stats}, we explore the identification and estimation of parameters in $\bXi_{m,n}$ and further investigate the rate of the order-adjusting parameter.

\section{Identifiability, estimation and statistical theory}\label{sec:stats}

 To streamline theoretical analysis and algorithmic implementation, we introduce a representation for hyperlinks using binary vectors. Specifically, for hyperlink $e_j$, we record $\by_{j}\in\{0,1\}^n$ where the $i$-th element $y_{ji}=\mathbf{1}_{\{ i\in e_j  \}}$. 
Denote $\balpha = (\alpha_1,\ldots,\alpha_n)^{\top}\in\RR^n$ and $\bZ = (\bz_1,\ldots,\bz_n)^{\top}\in\RR^{n\times K}$. 
Consider the logistic function applied component-wise to a vector $\ba\in\RR^n$ to yield $\sigma(\ba) = (\sigma(a_i), i \in [n])^{\top}\in[0,1]^n$.
Then we define $\bp_j =\sigma\big( \beta_{m,n}\mathbf{1}_n + \balpha + \bZ\bff_j\big)$, and we have that $\{\by_j\}_{j\in[m]}$ are conditionally independent observations from multivariate Bernoulli distributions:
\[
\by_j \mid \bXi_{m,n}  \sim \mathrm{multiBernoulli}(\bp_j),
\]
or equivalently, $y_{ji}$'s are conditionally independent Bernoulli random variables with means $ p_{ji}=\PP(y_{ji}=1\mid \bXi_{m,n}).$
We use the superscript $``*"$ to denote the true model parameters: $\bXi_{m,n}^* =  \big\{ \beta_{m,n}^*,\{\alpha_i^*\}_{i=1}^n,\{\bz_i^*\}_{i=1}^n,\{\bff_j^*\}_{j=1}^m  \big\}$. 
Denote $\balpha^*=(\alpha^*_1, \cdots, \alpha_n^*)^{\top}\in\RR^n$, $\bF^* = (\bff_1^*, \cdots, \bff_m^*)^{\top}\in\RR^{m\times K}$, $\bZ^* = (\bz_1^*, \cdots, \bz_n^*  )^{\top} \in\RR^{n\times K}$, and
$
 \bTheta^* = (\theta_{ji}^*)_{m\times n} = \beta^*_{m,n}\mathbf{1}_m\mathbf{1}_n^{\top} + \mathbf{1}_m\balpha^{*\top} + \bF^* \bZ^{*\top}\in\RR^{m\times n}$.  
{In the following, Section~\ref{sec:cmle} examines the identifiability of $\beta^*_{m,n}$ and $\balpha^*$. 
We study a constrained maximum likelihood estimator and establish its $F$-consistency in estimating 
$\bTheta^*$, $\balpha^*$, and $\beta^*_{m,n}$. 
Section~\ref{sec:pmle} further addresses the identification and estimation of all individual components in $\bXi_{m,n}^*$. 
We study another constrained maximum likelihood estimator with a more comprehensive set of constraints, show entry-wise consistency of the estimators, and derive their asymptotic distributions. 
These results further enable consistent estimation and asymptotic inference for each 
$\theta^*_{ji}$ and $p_{ji}^* = \sigma(\theta^*_{ji})$, for all $(j,i)\in[m]\times[n]$. 
}

\subsection{$F$-consistent  estimation for $\bTheta^*$, $\balpha^*$ and $\beta_{m,n}^*$}\label{sec:cmle}
Collect $\{\by_j\}_{j\in[m]}$ in a binary matrix such that  $\bY = (\by_1,\by_2,\ldots,\by_m)^{\top}\in\{0,1\}^{m\times n}$. 
The log-likelihood function is given by
$
   \cL(\beta, \balpha,\bF,\bZ | \bY ) := \sum_{j=1}^{m}\sum_{i=1}^n  \big\{y_{ji}\theta_{ji} - \log\big(1+\exp(\theta_{ji})\big)   \big\} $,
with $\bF=(\bff_1,\cdots,\bff_m)^{\top}$ and $\theta_{ji} = \beta+\alpha_i + \bff_j^{\top}\bz_i$. Consider the feasible region 
\beq\label{eq:constr_mle}
\begin{aligned}\cF_1(C_{\beta_{m,n}}):= \big\{ & (\beta, \balpha, \bF, \bZ)\in\RR\times \RR^{n} \times \RR^{m\times K} \times \RR^{n\times K}: \bJ_m\bF = \bF, \mathbf{1}_n^{\top}\balpha  = 0, \\[-0.5em]& \max_{i,j}\{|\alpha_i|, \|\bz_i \|_2 
 , \|\bff_j\|_2\}\le C_1,  -C_{\beta_{m,n}} \le \beta \le C_2  \big\},
\end{aligned}
\eeq
where $ C_1,C_2>0$  are some constants, and $\bJ_m=\bI_m - m^{-1}\mathbf{1}_m\mathbf{1}_m^{\top} $ is the centering operator.  
The boundedness constraint $ \max_{i,j}\{|\alpha_i|,  \|\bz_i \|_2, \|\bff_j\|_2 \}\le C_1$ is similar to those employed in parameter estimation for fitting dyadic network latent space models \citep[e.g.,][]{ma2020universal}.   Note that we allow $\beta_{m,n}^*$ to diverge towards $-\infty$ as $m,n\to\infty$, thus the parameter $C_{\beta_{m,n}}$ serves to control the rate of divergence of $\beta$ in the estimation, and the choice of the rate will be discussed in Remark \ref{rem:tune} below.

We first examine the statistical property of the CMLE (constrained maximum likelihood estimator) defined as follows:
\beq\label{eq:constr1_mle}
\begin{aligned}
    (\hat{\beta}, \hat{\balpha}, \hat{\bF}, \hat{\bZ}) = \max_{\beta, \balpha, \bF, \bZ} & \cL(\beta, \balpha,\bF,\bZ|\bY) ~~ \mathrm{s.t.} ~~ (\beta, \balpha, \bF, \bZ) \in \cF_1(C_{\beta_{m,n}}).
\end{aligned}
\eeq
The core of \eqref{eq:constr1_mle} is to find a low-rank estimator for $\bTheta^*$ based on its logistic-transformed, noisy observations,  and to recover $\beta_{m,n}^*$, $\balpha^*$ from the estimate for $\bTheta^*$.
For an estimator $\hat{\bTheta}\in\RR^{m\times n}$ of $\bTheta^*$, we call $\hat{\bTheta}$ an $F$-consistent estimator if $(mn)^{-\frac{1}{2}}\|\hat{\bTheta} - \bTheta^*\|_F = o_p(1)$ as $m,n\to\infty$. Theorem \ref{thm:samirt_mean_upper} below shows the $F$-consistency of the 
CMLE \eqref{eq:constr1_mle}.

\begin{thm}[$F$-consistent estimation of $\bTheta^*$]\label{thm:samirt_mean_upper} Let $\hat{\bTheta} = \hat{\beta}\mathbf{1}_m\mathbf{1}_n^{\top} + \mathbf{1}_m\hat{\balpha}^{\top} + \hat{\bF}\hat{\bZ}^{\top} $ where $   (\hat{\beta}, \hat{\balpha}, \hat{\bF}, \hat{\bZ}) $ is the solution to \eqref{eq:constr1_mle}.  For any $\epsilon>0$, as long as  $\exp(\beta^*_{m,n})\gtrsim \log\big((m\vee n)/\epsilon\big)/(m\wedge n)$ and $(\beta_{m,n}^*, \balpha^*,\bF^*,\bZ^* )\in\cF_1(C_{\beta_{m,n}}) $,  we have 
\beq\label{eq:thm31}
   \|\hat{\bTheta} - \bTheta^*\|_F \lesssim \frac{(m\vee n)^{\frac{1}{2}} \exp(\beta^*_{m,n}/2)\big\{ \log((m\vee n)/\epsilon\big\}^{\frac{1}{2}}}{\exp(-C_{\beta_{m,n}})},
\eeq
with probability at least $1-\epsilon$.
\end{thm}


\begin{rem} The condition
    $\exp(\beta^*_{m,n})\gtrsim \log((m\vee n)/\epsilon)/(m\wedge n) $ in Theorem \ref{thm:samirt_mean_upper}  suggests that as the number of vertices and the number of hyperlinks grow to infinity, the order of hyperlinks should be bounded below by a term proportional to $n\log((m\vee n)/\epsilon)/(m \wedge n)$. When $m\asymp n$, this order accords with the sufficient order in Proposition \ref{prop:suff_spar} and the necessary order in Proposition \ref{prop:nece_spar} up to a logarithmic factor. 
\end{rem}

\begin{rem}\label{rem:f_consist}
    Theorem \ref{thm:samirt_mean_upper} also suggests that if $m\asymp n$, $\exp(\beta^*_{m,n})\asymp \exp(-C_{\beta_{m,n}})$, and  $\exp(\beta^*_{m,n})\gg \log(m\vee n)/(m\wedge n)$, then the CMLE achieves $F$-consistent $\bTheta$ estimation, i.e., $(mn)^{-1/2}\|\hat{\bTheta} - \bTheta^*\|_F = o_p(1)$.
\end{rem}

\begin{rem}[A choice of $C_{\beta_{m,n}}$]\label{rem:tune}
 The choice of $C_{\beta_{m,n}}$ plays an important role in Theorem \ref{thm:samirt_mean_upper}. As shown in \eqref{eq:thm31}, a larger $C_{\beta_{m,n}}$ leads to a worse estimation error bound.   
 Meanwhile, $C_{\beta_{m,n}}$ needs to be large enough so that the condition $(\beta_{m,n}^*,\alpha^*,\bF^*,\bZ^*)\in\cF_1(C_{\beta_{m,n}})$ in Theorem \ref{thm:samirt_mean_upper} could hold.
 Such a $C_{\beta_{m,n}}$ can be set in practice via
     $C_{\beta_{m,n}} = C'\cdot\hat{C}_{m,n}$ where 
    \[
       \hat{C}_{m,n} = -\log\Big(\frac{\sum_{j=1}^m\sum_{i=1}^n \mathbf{1}_{\{y_{ji}=1\}}}{mn} \Big) 
    \]
    and $C'\ge 1$ is a constant that does not depend on $m,n$. In this construction, $\hat{C}_{m,n}$ offers a reliable estimate of $\beta_{m,n}^*$ and the constant $C'$ helps ensure that $\cF_1(C_{\beta_{m,n}})$ contains the true parameters.
    The following proposition states the effectiveness of this construction.
\end{rem}

\begin{prop}\label{prop:tune1}
    Suppose that there exists $s_{m,n}\to\infty$ such that $\exp(\beta_{m,n}^*)\gg s_{m,n}/(mn)$ as $m,n\to\infty$. Then the choice of $C_{\beta_{m,n}}$ in Remark \ref{rem:tune} satisfies $C_{\beta_{m,n}}\asymp -\beta^*_{m,n}$ with probability approaching $1$ as $m,n\to\infty$. Consequently, there exist constants $C'\ge 1$ in Remark \ref{rem:tune} and $C_1,C_2>0$ in \eqref{eq:constr_mle} such that 
    $
       \PP\big\{ (\beta_{m,n}^*, \balpha^*,\bF^*,\bZ^* )\in\cF_1(C_{\beta_{m,n}})  \big\} \to 1
    $
     as $m,n\to\infty$.
\end{prop} 

We also note that the $F$-consistency result in Theorem \ref{thm:samirt_mean_upper} leads to the following corollary, which  establishes $F$-consistency for the estimation of both ${\balpha}^*$ and $\beta^*_{m,n}$.

\begin{cor}[$F$-consistent estimation of $\beta_{m,n}^*$ and $\balpha^*$]\label{coro:beta_alpha}
    Let $   (\hat{\beta}, \hat{\balpha}, \hat{\bF}, \hat{\bZ}) $ be the solution to \eqref{eq:constr1_mle}.  For any $\epsilon>0$, as long as  $\exp(\beta^*_{m,n})\gtrsim \log\big((m\vee n)/\epsilon\big)/(m\wedge n)$ and $(\beta_{m,n}^*, \balpha^*,\bF^*,\bZ^* )\in\cF_1(C_{\beta_{m,n}}) $,  we have 
\[
\begin{aligned}
   &  \|\hat{\balpha} - \balpha^*\|_2 \lesssim \frac{(m\vee n)^{\frac{1}{2}} \exp(\beta^*_{m,n}/2)\big\{ \log((m\vee n)/\epsilon\big\}^{\frac{1}{2}}}{m^{\frac{1}{2}}\exp(-C_{\beta_{m,n}})},   \\
   & |\hat{\beta} - \beta_{m,n}^*|
 \lesssim \frac{(m\vee n)^{\frac{1}{2}} \exp(\beta^*_{m,n}/2)\big\{ \log((m\vee n)/\epsilon\big\}^{\frac{1}{2}}}{m^{\frac{1}{2}}n^{\frac{1}{2}}\exp(-C_{\beta_{m,n}})},
 \end{aligned}
\]
with probability at least $1-\epsilon$.
\end{cor}

Similar to Remark \ref{rem:f_consist}, if $m\asymp n$, $\exp(\beta^*_{m,n})\asymp \exp(-C_{\beta_{m,n}})$, and  $\exp(\beta^*_{m,n})\gg \log(m\vee n)/(m\wedge n)$, we obtain $F$-consistent estimation of both $\balpha^*$ and $\beta^*_{m,n}$, i.e., $n^{-1/2} \|\hat{\balpha} - \balpha^*\|_2 = o_p(1) $ and $|\hat{\beta} - \beta_{m,n}^*| = o_p(1)$.

\subsection{Entry-wise consistent estimation and asymptotic inference for embedding vectors}\label{sec:pmle}
In this section, we establish entry-wise consistent estimation and asymptotic inference for all the individual components in $\bXi_{m,n}^*$. 
To make the embedding vectors identiafiable, further assumptions need to be made.
The identification issue is due to the fact that for any 
full rank $\bA\in\RR^{K\times K}$, letting 
$\tilde{\bff}_j = \bA^{\top}\bff_j^*$ and $\tilde{\bz}_i = \bA^{-1}\bz_i^*$ leads to 
\beq\label{eq:equiv}
\begin{aligned}
  \tilde{\bff}_j^{\top}\tilde{\bz}_i =   \bff_j^{*\top}\bA\bA^{-1} \bz^*_i 
 =  {\bff}_j^{*\top}\bz_i^*,
\end{aligned}
\eeq
which preserves the likelihood in \eqref{eq:hyper_model}, making individual embeddings not identifiable from observed data. To address the identifiability issue,
we assume a set of embedding vector constraints as follows:
$$
\text{(C1) Both}~ \bF^{*\top}\bF^{*}~\text{and}~ \bZ^{*\top}\bZ^*~\text{are diagonal;} ~~
    \text{(C2)}~ \frac{\bF^{*\top}\bF^*}{m} = \frac{\bZ^{*\top}\bZ^*}{n};~~
    \text{(C3)}~ \bJ_m\bF^* = \bF^*. 
$$
The third constraint is from Section \ref{sec:model} for the identification of vertex degree parameters and the first two are for identifying the linear transformation $\bA$.
If the true embedding vectors do not satisfy the first two constraints, a transformation of the embeddings that satisfies the constraints can be performed as follows.   
Let $\cV = \diag( \rho_1^2,\cdots,\rho_K^2 )$ with $\rho_1^2>\rho_2^2>\cdots >\rho_K^2$ being the eigenvalues of $(mn)^{-1}(\bZ^{*\top}\bZ^*)^{1/2}\bF^{*\top}\bF^*(\bZ^{*\top}\bZ^*)^{1/2}$ and let $\bGamma$ be a $K\times K$ matrix whose columns collect the corresponding eigenvectors. Let $\bG = \big( \bZ^{*\top}\bZ^*/n \big)^{1/2}\bGamma\cV^{-\frac{1}{4}}$. The transformed $(\bZ^*, \bF^*)$ is $
\big( \bZ^*(\bG^{\top})^{-1}, \bF^*\bG \big)
$.
One interpretation of this transformation is that we fix 
$\bA = \bG$ in \eqref{eq:equiv}. 
  Another identifiability issue concerns with column sign flip. 
   In particular, let $\bD$ be a $K\times K$ diagonal matrix with $\{-1,1\}$ in its  diagonal entries. Note  that $\bD$ is symmetric and $\bD^2 = \bI$. Even when $\bF^*$ and $\bZ^*$ satisfy the above embedding vector constraints, if we  let $\tilde{\bF} = \bF^* \bD$ and $\tilde{\bZ} = \bZ^* \bD$, $\tilde{\bF}$ and $\tilde{\bZ}$ would still satisfy the constraints and it holds that  $\tilde{\bF}\tilde{\bZ}^{\top} =\bF^*\bD\bD\bZ^{*\top} = \bF^*\bZ^{*\top}$.  Consequently, $\bF^*$ and $\bZ^*$ are identifiable up to column sign flips.  We deal with this issue by fixing the sign of all coordinates in the first vertex embedding vector and adjust the estimators accordingly. In comparison to the identification of the linear transformation, the identification issue with column sign flipping is 
   minor.



Under the above identifiability conditions, we establish entry-wise consistency and asymptotic normality in the estimation of all individual parameters in $\bXi_{m,n}$.   Note that due to the constraint $\sum_{i\in[n]}\alpha^*_i=0$, there are indeed $n$ rather than $n+1$ free parameters in $(\beta_{m,n}^*,\balpha^*)$. 
For convenience in the theoretical analysis, we re-parameterize the model by integrating  $\beta_{m,n}^*$ and $\balpha^*$ into an $\balpha^\dagger$ term, which results in un-centered vertex degree parameters. Specifically, we let $\alpha_i^\dagger = \beta_{m,n}^* + \alpha_i^*$ for $i\in[n]$. We now aim to estimate individual parameters in $(\balpha^\dagger,\bF^*,\bZ^*)$. 

Let $\bff = (\bff_1^\top,\ldots,\bff_{m}^\top)^{\top}$ and $\bz = (\bz_1^\top,\ldots,\bz_{n}^{\top})^{\top}$ denote vectorized versions of $\bF$ and $\bZ$, where $\bff_j^\top\in\RR^K$ is the $j$th row of $\bF$ and $\bz_i^\top\in\RR^K$ is the $i$th row of $\bZ$. Similarly, let $\bff^*$ and $\bz^*$ denote vectorized versions of $\bF^*$ and $\bZ^*$, respectively. Under the new parametrization, we rewrite the feasible region $\cF_1(C_{\beta_{m,n}})$ in \eqref{eq:constr_mle} in the following form: 
\beq\label{eq:feasible_entry}
\begin{aligned}   \cF_2(C_{\beta_{m,n}}):= \{ &(\balpha,\bz,\bff)\in \RR^{n}\times \RR^{nk}\times \RR^{mk}: -C_{\beta_{m,n}} \le \sum_{i=1}^{n} \alpha_i / n \le C_3 , \\
    & \|\balpha -  \sum_{i=1}^{n} \alpha_i / n \cdot \mathbf{1}_n \|_\infty \le C_4, ~  \max\{\max_{j\in[m]}\|\bff_j\|_{2},  \max_{i\in[n]}\|\bz_i\|_{2}\}\le C_5, \\ &  \text{ where } C_3 \text{ is specified below and }C_4,C_5>0 \text{ are constants}.  \}. 
\end{aligned}
\eeq

For the technical derivation of entry-wise consistent results and asymptotic distributions, we further set $C_3 = - C_3'\cdot C_{\beta_{m,n}} $ for some $C_3'\in(0,1)$. 
As shown in the following proposition, this choice of $C_3$ retains the good properties of $\cF_1(C_{\beta_{m,n}})$, where $C_{\beta_{m,n}}$ is taken as in Remark \ref{rem:tune}.   
{The additional constraint for $C_3$ is mainly for technical purposes, and practically in algorithmic implementation, we find that the choice of $C_3$ can be flexible and does not significantly change the results.}
\begin{prop}\label{prop:tune2}
    Following the conditions in Proposition \ref{prop:tune1}, 
    there exist constants $C'\ge 1$ in Remark \ref{rem:tune} and $C_3'\in(0,1),C_4,C_5>0$ such that
    $
       \PP\big\{ (\balpha^\dagger,\bff^*,\bz^* )\in\cF_2(C_{\beta_{m,n}})  \big\} \to 1
    $
     as $m,n\to\infty$.
\end{prop}

We also re-write the log-likelihood function as
$
\cL(\bY |\balpha, \bz,\bff) = \sum_{j=1}^m\sum_{i=1}^n\big\{ y_{ji}(\alpha_i+\bff_j^{\top} \bz_i) - \log\big( 1+ \exp( \alpha_i+\bff^{\top}_j\bz_i )   \big)    \big\}$. 
We consider the following constrained maximum likelihood estimator
\beq\label{eq:cmle2}
\begin{aligned}
    \max_{\balpha,\bz,\bff} \cL(\bY|\balpha,\bz,\bff) ~~\text{s.t.}~~ & \frac{1}{n}\bZ^{\top}\bZ= \frac{1}{m}\bF^{\top}\bF,      ~\bZ^{\top}\bZ  \text{ and } \bF^{\top}\bF \text{ are diagonal}, \\ &
   \bJ_m \bF = \bF, (\balpha,\bz,\bff) \in \cF_2(C_{\beta_{m,n}}). 
\end{aligned}
\eeq
{Under the following conditions on the embedding vectors $\{\bff_j^*\}_{j\in[m]}$ and $\{\bz_i^*\}_{i\in[n]}$, the solution to \eqref{eq:cmle2} can achieve $F$-consistent estimation for $\{\balpha^\dagger,\bz^{*},\bff^{*}\}$. The $F$-consistency results will later be used for deriving entry-wise consistency results and asymptotic distributions.}


\begin{cond}[Eigen-structures and norm bounds of the embedding vectors]\label{assum:1}
We assume 
\beq\label{eq:cond1}
\begin{aligned}
    \rho_{\min}(m^{-1}\sum_{j=1}^{m}\bff^{*}_j\bff_j^{*\top}) = \rho_{\min}(n^{-1}\sum_{i=1}^{n}\bz^{*}_i\bz_i^{*\top}) \gtrsim 1, 
\end{aligned}
\eeq
  $\max_{j\in[m]}\|\bff_j^*\|_{\infty}\lesssim 1$, and $\max_{i\in[n]}\|\bz_i^*\|_{\infty}\lesssim 1$. Moreover,  the eigenvalues of each of the $K\times K$ matrices
$m^{-1}\sum_{j=1}^m \bff^{*}_j\bff^{*\top}_j, n^{-1} \sum_{i=1}^n \bz_i^{*}\bz_i^{*\top}${ and }$(m^{-1}\sum_{j=1}^m \bff^{*}_j\bff^{*\top}_j) (n^{-1} \sum_{i=1}^n \bz_i^{*}\bz_i^{*\top})$ are distinct and their gaps are lower bounded by a constant.
\end{cond}

Condition \ref{assum:1} ensures that the embedding vectors' entries remain finite, same as the assumptions made in Section \ref{sec:model}. Additionally, it requires  non-degenerate second moment structures, thereby ensuring that each of the $K$ dimensions is non-trivial. The equal sign in \eqref{eq:cond1} comes from the identifiability conditions. The distinctness assumption aims to ensure unique identification of the columns of $\bF^*$ and $\bZ^*$. Under these regular conditions, the $F$-consistency of the estimators in \eqref{eq:cmle2} are established as follows.

\begin{prop}[$F$-consistent estimation for $(\balpha^\dagger, \bz^*, \bff^*)$]\label{prop:average} Let $ (\hat{\balpha}, \hat{\bz},\hat{\bff})$ be the solution to \eqref{eq:cmle2}. Under Condition \ref{assum:1}, assume that  $(\balpha^\dagger, \bz^*, \bff^*)\in\cF_2(C_{\beta_{m,n}})$.  Define 
\beq\label{eq:delta_mn}
 \delta_{m,n} = \frac{(m\vee n)^{\frac{1}{2}} \exp(\beta_{m,n}^*/2)\big\{ \log((m\vee n)\big\}^{\frac{1}{2}}}{m^{\frac{1}{2}}n^{\frac{1}{2}}\exp(-C_{\beta_{m,n}})}.
\eeq
Then as $(m,n)\to\infty$, we have 
\[
 \begin{aligned}
       & n^{-\frac{1}{2}}\| \hat{\balpha} - \balpha^{\dagger} \|_2 = O_p ( \delta_{m,n}), 
       ~~ n^{-\frac{1}{2}}\| \hat{\bz} - \bz^{*} \|_2 = O_p (\delta_{m,n}), 
       ~~ m^{-\frac{1}{2}}\| \hat{\bff} - \bff^{*} \|_2 =  O_p (\delta_{m,n}).
  \end{aligned}
\]
\end{prop}

\begin{rem}
    From Proposition \ref{prop:average}, if $m\asymp n$, $\exp(\beta_{m,n}^*)\asymp \exp(-C_{\beta_{m,n}})$, $\exp(\beta_{m,n}^*)\gg \log(m\vee n)/(m\wedge n)$, we achieve the $F$-consistent estimation for $\balpha^\dagger$, $\bz^*$ and $\bff^*$, i.e., $n^{-1/2} \|\hat{\balpha} - \balpha^\dagger\|_2 = o_p(1) $, $(nK)^{-1/2}\|\hat{\bz} - \bz^* \|_2 = o_p(1)$ and $(mK)^{-1/2}\|\hat{\bff} - \bff^*\|_2 = o_p(1)$. 
\end{rem}

{
Next, we study entry-wise consistency properties and asymptotic normality of the estimators.}
This analysis presents several significant challenges.  
First, the  minimum  eigenvalue  of the Hessian of the log-likelihood function  is intricate to track due to identifiability issues and non-linearity of the problem. Moreover, the sparsity introduced by the diverging parameter $\beta_{m,n}^*$ as $m,n\to\infty$ further complicates the problem.
 To address these theoretical challenges, we introduce the following penalized and constrained maximum likelihood problem by using the method of Lagrange multipliers: 
\beq\label{eq:pmle2}
\begin{aligned}
\max_{\balpha,\bz,\bff} \cQ(\balpha,\bz,\bff|  \lambda_{m,n}) & = \cL(\bY|\balpha, \bz,\bff) - \cP(\bz,\bff|\lambda_{m,n}), \text{ s.t. }  (\balpha,\bz,\bff) \in \cF_2(C_{\beta_{m,n}}), \\
 \mbox{with }    \cP(\bz,\bff|\lambda_{m,n})  = & ~ \frac{\lambda_{m,n} mn}{8}\Big\| \text{diag}\Big(\frac{1}{n}\bZ^{\top}\bZ - \frac{1}{m}\bF^{\top}\bF \Big)  \Big\|_F^2 + \frac{\lambda_{m,n} mn  }{2}\Big\|\frac{\bF^{\top}\mathbf{1}_m}{m} \Big\|_2^2  \\ & ~ +\frac{\lambda_{m,n} mn}{2} \Big\| \text{ndiag}\Big( \frac{1}{m}\bF^{\top}\bF  \Big) \Big\|_F^2 + \frac{\lambda_{m,n} mn}{2} \Big\| \text{ndiag}\Big( \frac{1}{n}\bZ^{\top}\bZ  \Big) \Big\|_F^2,
\end{aligned}
\eeq
where $\lambda_{m,n}>0$ is the Lagrange multiplier parameter. For a squared matrix $\bA$, $\text{diag}(\bA)$ stands for the diagonal matrix with diagonal entries equal to those of $\bA$ and $\text{ndiag}(\bA)$ is the matrix containing upper triangular entries of $\bA$, 
i.e., the $(i,j)$-th entry of $\text{ndiag}(\bA)$ is zero if $i\le j$ and $A_{ij}$ if $i>j$.  
The penalty   $\cP(\bz,\bff|\lambda_{m,n})$ in \eqref{eq:pmle2} is based on identifiability constraints in \eqref{eq:cmle2}. The next proposition shows the equivalence between \eqref{eq:pmle2} and \eqref{eq:cmle2}.
\begin{prop}\label{prop:equiv}
    For any $\lambda_{m,n}>0$,  the solutions to \eqref{eq:cmle2} and \eqref{eq:pmle2} are equivalent.
\end{prop}

{
Since the solution remains unchanged for any $\lambda_{m,n}>0$, we can take advantage of the penalization term   $ \cP(\bz,\bff|\lambda_{m,n})$ in \eqref{eq:pmle2} when analyzing \eqref{eq:cmle2}: in combination with the Hessian matrix of the log-likelihood function, the Hessian of the penalty function $\cP$ expands the entire linear space $\RR^{n(K+1)+mK}$ with high probability.
Similar Lagrangian multiplier strategies have been used in the literature, such as \cite{silvey1959lagrangian} and \cite{el1994wald} for Lagrangian multiplier test and \cite{wang2022maximum} for generalized factor models with application to factor-augmented regressions. 
Due to space constraints, 
detailed theoretical analyses are in the supplementary material and we summarize the main results below.

We begin by introducing the following two conditions.
}
 
\begin{cond}[Hyperlink order]\label{assum:3}
    There exists a small $\epsilon^*>0$ such that  $ \exp(\beta_{m,n}^*) \gg (m\vee n)^{\epsilon^*}/(m\wedge n) $, as $m,n\to\infty$. 
\end{cond}

\begin{cond}[Constraint parameter in \eqref{eq:feasible_entry}]\label{assum:2}
    $C_{\beta_{m,n}} \asymp  |\beta^*_{m,n}|$, as $m,n\to\infty$.
\end{cond}

Condition \ref{assum:3} requires a minimum order for the hyperlinks which accords with the discussion in Section \ref{sec:first_gl}. Condition \ref{assum:2} indicates that the choice of $C_{\beta_{m,n}}$ should be reasonable based on the order-adjusting parameter.

\begin{thm}[{Consistency of the embedding vectors and un-centered vertex degrees estimation}]\label{thm:entry-wise-consistency}
 Let $ (\hat{\balpha}, \hat{\bz},\hat{\bff})$ be the solution to \eqref{eq:cmle2}.
 Under Conditions \ref{assum:1}, \ref{assum:3} and \ref{assum:2}, assume that  $(\balpha^\dagger, \bz^*, \bff^*)\in\cF_2(C_{\beta_{m,n}})$. Then we have
 \[
  \begin{aligned} 
   \max\big\{\|\hat{\balpha}-\balpha^\dagger\|_{\infty},\|\hat{\bz}-\bz^*\|_{\infty},\|\hat{\bff}-\bff^*\|_{\infty}  \big\}   = O_p\Big\{\frac{(m+n)^{\epsilon}}{(m\wedge n)^{\frac{1}{2}}\exp(\beta_{m,n}^*/2) } \Big\}
  \end{aligned}
 \]
for any $\epsilon>0$, as $(m,n)\to\infty$. {Moreover, we have an improved average consistency rate as}
\beq\label{eq:thm:imp:ave}
    {\frac{1}{m}\sum_{j=1}^m\|\hat{\bff}_j - \bff_j^*\|_2^2 + \frac{1}{n}\sum_{i=1}^n\|\hat{\bz}_i - \bz_i^*\|_2^2 + \frac{1}{n}\|\hat{\balpha}-\balpha^\dagger\|_2^2 = O_p\Big\{ \frac{1}{(m\wedge  n)\exp(\beta_{m,n}^*)} \Big\}.}
\eeq
\end{thm}
{The improved average consistency rate in \eqref{eq:thm:imp:ave} eliminates the $\log(m \vee n)$ term in Proposition~\ref{prop:average}. This improvement is from an intermediate result in the proof of uniform consistency  where we show that the score function of $\cQ(\balpha,\bz,\bff| \lambda_{m,n})$ in \eqref{eq:pmle2} evaluated at $(\hat{\balpha}, \hat{\bz}, \hat{\bff})$ equals $\mathbf{0}$ with high probability.}
Let $\bP^* = (p^*_{ji})_{m\times n} = \sigma(\bTheta^*)$. 
The next corollary 
demonstrates entry-wise consistent estimation of $\theta^*_{ji}$ and $p_{ji}^*$ based on \eqref{eq:cmle2}.

\begin{cor}[Consistency of $\bTheta$ and $\bP$ estimation]\label{cor:natural}
    Let $ (\hat{\balpha},\hat{\bz},\hat{\bff})$ be the solution to \eqref{eq:cmle2}, $\hat{\bTheta} = (\hat{\theta}_{ji})_{m\times n}$ with $\hat{\theta}_{ji} =\hat{\alpha}_i + \hat{\bff}_{j}^{\top} \hat{\bz}_{i}$,  and $\hat{\bP} = \sigma(\hat{\bTheta})$. Then under the conditions in Theorem \ref{thm:entry-wise-consistency}, for any $\epsilon>0$, we have
    \[
    \begin{aligned}
        & \|\hat{\bTheta} - \bTheta^* \|_{\max} =  O_p\Big\{\frac{(m+n)^{\epsilon}}{(m\wedge n)^{\frac{1}{2}}\exp(\beta_{m,n}^*/2) } \Big\}  \text{ and } \\ & 
        \|\hat{\bP} - \bP^* \|_{\max} =  O_p\Big\{\frac{(m+n)^{\epsilon}\exp(\beta_{m,n}^*/2)}{(m\wedge n)^{\frac{1}{2}} } \Big\}
    \end{aligned}
    \] 
    as $(m,n)\to\infty$.
\end{cor}



In the following, we present asymptotic distributions for the estimators in \eqref{eq:cmle2}. 
{Let $\bnu^*_i = (\alpha_i^{\dagger}, \bz_i^{*\top})^{\top}$ for $i \in [n]$. The following theorem presents the individual and joint asymptotic distributions of $(\hat{\alpha}_i, \hat{\bz}_i^{\top})^{\top}$ and $\hat{\bff}_j$.
}

{
\begin{thm}[Latent embedding asymptotic distributions]\label{thm:limit_dist}
     Let $ (\hat{\balpha}, \hat{\bz},\hat{\bff})$ be the solution to \eqref{eq:cmle2} and $\hat{\bnu}_i = (\hat{\alpha}_i,\hat{\bz}_i^{\top})^{\top}$. Under the conditions in Theorem \ref{thm:entry-wise-consistency} and further assume that $\exp(\beta_{m,n}^*/2)\gg (m\vee n)^{\frac{1}{2}+\epsilon^*} /(m\wedge n) $ as $m,n\to\infty$. We have
     \begin{itemize}
         \item \emph{(Individual)} $\exp(\beta_{m,n}^*/2)m^{\frac{1}{2}}\bSigma^{-\frac{1}{2}}_{(\bnu_i,m,n)}(\hat{\bnu}_i-\bnu_i^*) \overset{ d}{\rightarrow} \cN\big(\mathbf{0},    \bI\big)$ and \\ $\exp(\beta_{m,n}^*/2)n^{\frac{1}{2}}\bSigma_{(\bff_j,m,n)}^{-\frac{1}{2}}\big( \hat{\bff}_j - \bff_j^{*} \big) \overset{ d}{\rightarrow} \cN\big(\mathbf{0},    \bI \big)$;
         \item \emph{(Joint)} $ \exp(\beta_{m,n}^*/2)\bSigma_{(\bnu_i,\bff_j,m,n)}^{-\frac{1}{2}}\big\{ m^{\frac{1}{2}}(\hat{\bnu}_i - \bnu_i^*)^{\top}, n^{\frac{1}{2}}(\hat{\bff}_j - \bff_j^*)^{\top}\big\}^{\top}  \overset{ d}{\longrightarrow}  \cN(\mathbf{0}_{2K+1}, \bI_{2K+1}) $, 
     \end{itemize}
    as $m,n\to\infty$, where $\bSigma_{(\bnu_i,m,n)}$, $\bSigma_{(\bff_j,m,n)}$, and $\bSigma_{(\bnu_i,\bff_j,m,n)}$ denote the asymptotic variance terms, 
with their full definitions provided in Section~4.3.1 of the supplementary material.

\end{thm}
}

Building upon the joint limit distributions in Theorem \ref{thm:limit_dist}, we present limit distributions for estimates of $\theta_{ji}^*$ and $p_{ji}^*$ in Corollary \ref{coro:theta_p}. 

{
\begin{cor}[Asymptotic distributions for $\theta_{ji}^*$ and $p_{ji}^*$ estimation]\label{coro:theta_p} 
 Let $ (\hat{\balpha}, \hat{\bz},\hat{\bff})$ be the solution to \eqref{eq:cmle2}. Let $\hat{\theta}_{ji} = \hat{\alpha}_i + \hat{\bff}_j^{\top}\hat{\bz}_i$ and $\hat{p}_{ji} = \sigma(\hat{\theta}_{ji})$. Under the conditions in Theorem \ref{thm:limit_dist}, we have
    \beq\label{eq:limit_theta}
        \begin{aligned}
        \exp(\beta_{m,n}^*/2) (m \wedge n )^{\frac{1}{2}}\mathrm{Var}_{m,n}^{-\frac{1}{2}}{(\theta_{ji})}(\hat{\theta}_{ji} - \theta_{ji}^*)  \overset{ d}{\longrightarrow} \cN(0, 1 ),
        \end{aligned}
    \eeq
      \beq\label{eq:limit_p}
        \begin{aligned}
            \exp(- \beta_{m,n}^*/&2)  (m \wedge n )^{\frac{1}{2}}\mathrm{Var}^{-\frac{1}{2}}_{m,n}{(p_{ji})}(\hat{p}_{ji} - p_{ji}^*)  \overset{ d}{\longrightarrow} \cN(0, 1 ),
        \end{aligned}
    \eeq
    as $m,n\to\infty$, where $\mathrm{Var}_{m,n}{(\theta_{ji})}$ and $\mathrm{Var}_{m,n}{(p_{ji})}$ are asymptotic variance terms with their full definitions provided in Section~4.3.1 of the supplementary material.
\end{cor}
}


Theorem \ref{thm:limit_dist} and Corollary \ref{coro:theta_p} provide the basis for constructing confidence intervals for $(\alpha_i^\dagger, \bz_i^*, \bff_{j}^*)$, $\theta_{ji}^*$ and $p_{ji}^*$. 
To complete the inference task, we plug in the estimators $(\hat{\balpha}, \hat{\bz},\hat{\bff})$ into the covariance expressions. 
Let $\wh{\bSigma}_{(\bnu_i,m,n)}$ and $\wh{\bSigma}_{(\bff_j,m,n)}$ denote the plug-in estimators for ${\bSigma}_{(\bnu_i,m,n)}$ and ${\bSigma}_{(\bff_j,m,n)}$, respectively. Let $\wh{\mathrm{Var}}_{m,n}{(\theta_{ji})}$ and $\wh{\mathrm{Var}}_{m,n}{(p_{ji})}$ be the plug-in estimators for ${\mathrm{Var}}_{m,n}{(\theta_{ji})}$ and ${\mathrm{Var}}_{m,n}{(p_{ji})}$, respectively. The following corollary establishes the consistency of the plug-in estimators.

\begin{cor}\label{coro:var_limit} Let $\delta'_{m,n,\epsilon} = {(m+n)^{\epsilon}}{(m\wedge n)^{-\frac{1}{2}}\exp(-\beta_{m,n}^*/2)} $. Under the conditions in Theorem \ref{thm:limit_dist}, the plug-in estimators of the variance-covariance matrices satisfy that
\begin{itemize}
    \item $\|\wh{\bSigma}_{(\bnu_j,m,n)} -  \bSigma_{(\bnu_j,m,n)}\|_F = O_p(\delta'_{m,n,\epsilon})$, $\|\wh{\bSigma}_{\bff_j,m,n} -  \bSigma_{\bff_j,m,n}\|_F = O_p(\delta'_{m,n,\epsilon})$;
    \item $ |\wh{\mathrm{Var}}_{m,n}{(\theta_{ji})} - {\mathrm{Var}}_{m,n}{(\theta_{ji})} | = O_p(\delta'_{m,n,\epsilon}) $, $ |\wh{\mathrm{Var}}_{m,n}{(p_{ji})} - {\mathrm{Var}}_{m,n}{(p_{ji})}  | = O_p(\delta'_{m,n,\epsilon}) $,
\end{itemize}
for any $\epsilon>0$
as $m,n\to \infty$. 
\end{cor}

Combining Theorem \ref{thm:limit_dist}, Corollaries \ref{coro:theta_p} and \ref{coro:var_limit} and applying Slutsky's Theorem, we have that the asymptotic normality results hold for the plug-in variance-covariance terms.

\section{Simulation studies}\label{sec:numerical}

In this section, we conduct simulation studies to evaluate the performance of the proposed method and validate our theoretical findings. 
To compute the estimators, we employ a projected gradient descent algorithm \citep{ma2020universal} and use a universal singular value thresholding  strategy \citep{chatterjee2015matrix}  for initialization. Details of the algorithms are in the supplementary materials. 
We consider various simulation setups where the embedding vectors are generated from a mixture of truncated multivariate Gaussian distributions. Specifically, when considering a $K$-dimensional latent space, we first randomly split the $n$ vertices into $K$ disjoint almost equal-sized groups, where ``almost" means the sizes can differ by $1$ when $K$ does not divide $n$. Vertex embeddings are then independently generated as follows. For vertex $i$, if it is in the $k$th group, $\bz^*_i$ would be generated from $\cN_{[-1,1]}(\be_k, \bSigma_{\bz} )$, where  $\be_k\in\{0,1\}^K$ is a column vector with the $k$th element being $1$ while other entries being $0$, $\bSigma_{\bz}$ is the covariance matrix of the embedding vectors, and $[-1,1]$ indicates truncation around the mean by deviation 1 on each coordinate. The embedding vectors $\{\bff^*_j\}_{j\in[m]}$ of the hyperlinks are centered i.i.d.  realizations of $\cN_{[-1,1]}(\mathbf{0}, \bSigma_{\bff})$ so that $\sum_{j\in[m]} \bff^*_j = \mathbf{0}$. For both $\bSigma_{\bz}$ and $\bSigma_{\bff}$, we adopt the autoregressive design where $\Sigma_{\bz,ij} = \Sigma_{\bff,ij} =0.2* \rho^{|i-j|}$ with $\rho\in\{0,0.5\}$. In particular, $\rho =0$ gives independent dimensions of the embedding vectors.  The vertex degree heterogeneity parameters $\{\alpha_i^*\}\subset \RR$ are i.i.d. realizations of $\mathrm{Uniform}[-1, 1]$ and the order-adjusting parameter $\beta^*_{m,n}\in\RR$ will be specified in each simulation setup. The hyperlinks $\by_{j}$'s are generated independently via 
$
   \by_{j} \sim  \mathrm{multiBernoulli}\big\{ \sigma ( \btheta_j^* )   \big\}$,
where $\btheta^*_j = \beta^*_{m,n}\mathbf{1}_n + \balpha^* + \bZ^*\bff^*_j $.
In Section \ref{sec:simu_theta},
we investigate the effect of sample size, the order-adjusting parameter and the dimension of the latent space on the estimation of $\bTheta^* = (\btheta^*_1,\ldots.\btheta^*_m)^{\top}\in\RR^{m\times n}$.  
In Section \ref{sec:simu_limit}, we construct asymptotic confidence intervals for the individual parameters $(\balpha^\dagger, \bF^\bG, \bZ^\bG)$ and $(\bTheta^*,\bP^*)$ based on the asymptotic normality results in Section \ref{sec:pmle},  and evaluate their finite sample performance. 

\subsection{$\bTheta$ estimation}\label{sec:simu_theta}

\begin{figure}[t]
      \centering
      \setlength\tabcolsep{0pt}
	\renewcommand{\arraystretch}{-5}      
      \begin{tabular}{cc}
      \includegraphics[width=7.5cm]{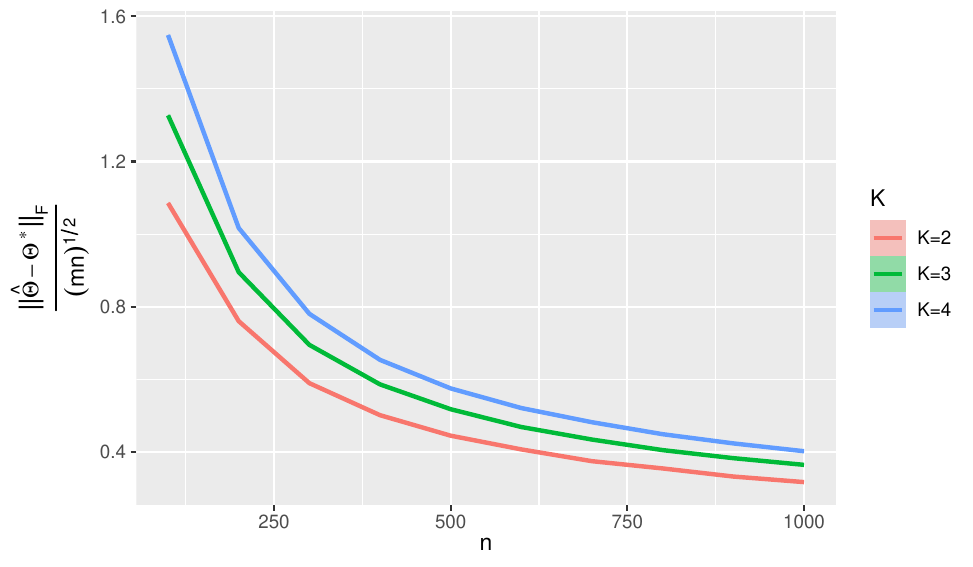} & \includegraphics[width=7.5cm]{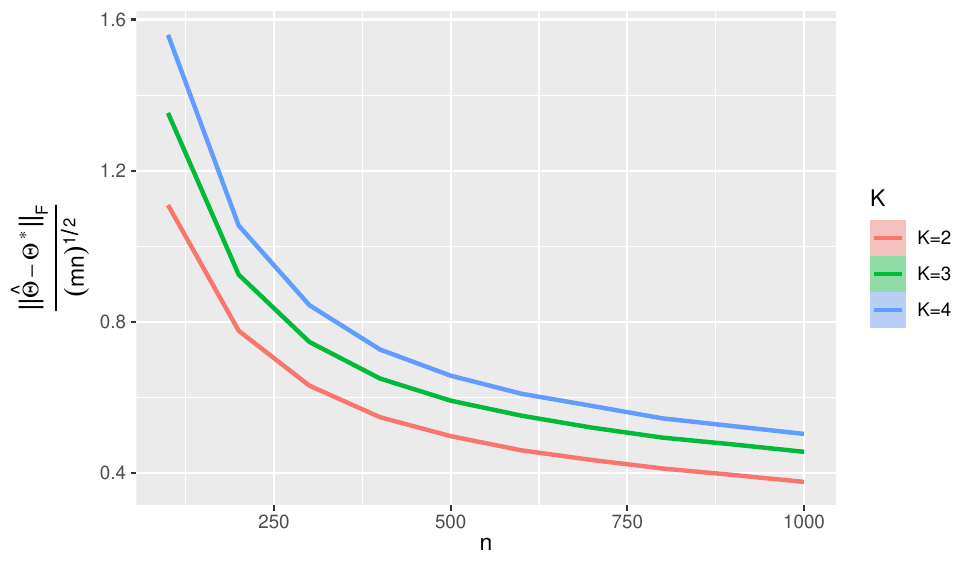}  \\  
    (a) $\rho = 0$ & (b) $\rho = 0.5$ \\
      \end{tabular}
      \caption{\footnotesize{Estimation error versus sample size based on $100$ Monte Carlo  repetitions.}}
      \label{fig:theta_mn}
\end{figure}

We first evaluate the effect of sample size and the dimension of the latent space together. We let $K \in\{2,3,4\}$, $\beta_{m,n}^* = -3$ and set $m = 10n$ with $n\in\{100,200,\cdots, 1000\}$. Figure \ref{fig:theta_mn} shows how the sample size changes the estimation error, with different setups of the latent space dimension.
We can observe that as $(m,n)$ grows, the relative Frobenius error decreases proportionally to $(m\wedge n)^{-1/2} = n^{-1/2}$ in shape, which is consistent with the estimation error bound  of Theorem \ref{thm:samirt_mean_upper}. As the dimension of the latent embedding space grows, the estimation error grows correspondingly. Though not displayed in the error bound in Theorem \ref{thm:samirt_mean_upper} since we consider non-diverging latent space dimension in the theoretical analysis, an increase in $K$ would result in the increase of estimation error, due to the proof of Theorem \ref{thm:samirt_mean_upper} in the supplementary material. With higher correlations among dimensions of the embedding vectors, the estimation error remains almost the same and gets slightly higher as the dimension grows.

\begin{figure}[t]
      \centering
      \setlength\tabcolsep{0pt}
	\renewcommand{\arraystretch}{-5}      
      \begin{tabular}{cc}
      \includegraphics[width=7.5cm]{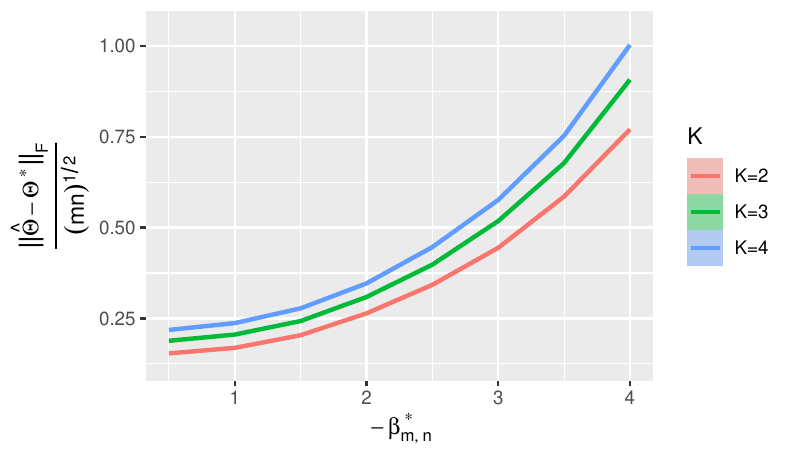} & \includegraphics[width=7.5cm]{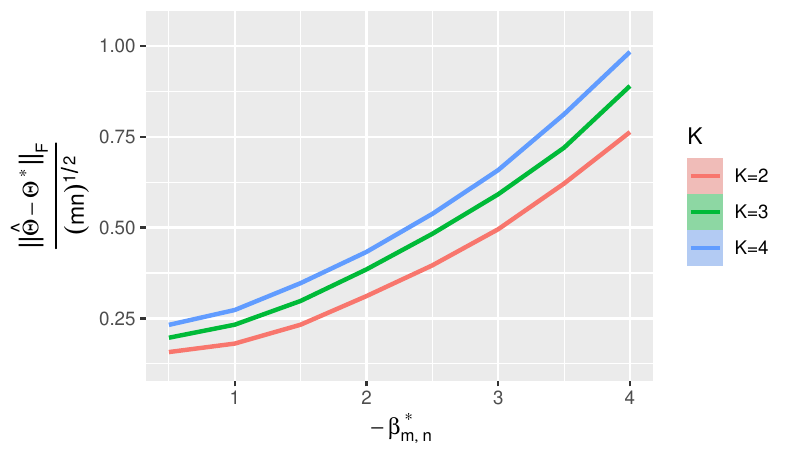}  \\  
    (a) $\rho = 0$ & (b) $\rho = 0.5$ \\
      \end{tabular}
      \caption{\footnotesize{Estimation error  versus the order-adjusting parameter based on $100$ Monte Carlo repetitions.}}
      \label{fig:theta_beta}
\end{figure}

We further investigate the effect of the order-adjusting parameter under different latent space dimensions. We let $K \in \{2,3,4\}$, $m =10n= 5,000 $ and set $-\beta^{*}_{m,n}\in\{0.5,1,\cdots, 3.5, 4\}$. 
The results are displayed in Figure \ref{fig:theta_beta}. 
We can see that as the value of $-\beta^*_{m,n}$ gets bigger, the estimation error grows proportional to $\exp(-\beta^*_{m,n})$.
 The effects of the dimension of the latent embedding space $K$ and the correlation structure among various dimensions are similar to those in Figure \ref{fig:theta_mn}.

Note that for conciseness, we focus on the presentation of the estimation error of $\bTheta^*$  in this subsection, while omitting the results for $\balpha^*$ and $\beta_{m,n}^*$. 
The estimation errors for $\balpha^*$ and $\beta_{m,n}^*$ are highly consistent with   $\|\hat{\bTheta} - \bTheta^* \|_F$ by noticing that $\hat{\balpha} - \balpha^* = m^{-1}\mathbf{1}_{m}\mathbf{1}_{m}^{\top}(\hat{\bTheta} - \bTheta^*)$ and $\hat{\beta} - \beta^*_{m,n} = n^{-1}\mathbf{1}_{n}^{\top}(\hat{\balpha} - \balpha^*  )$. 

\subsection{Confidence intervals}\label{sec:simu_limit}

\begin{figure}[t]
    \centering
      \setlength\tabcolsep{0pt}
	\renewcommand{\arraystretch}{-5}      
      \begin{tabular}{cc}   
      \multicolumn{2}{c}{\includegraphics[width = 8.2cm]{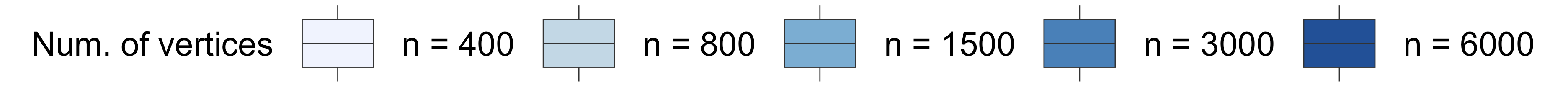}  } \vspace*{0.3cm} \\
      ~ & ~ \\ \vspace*{0.5cm}
      \includegraphics[width=7.2cm]{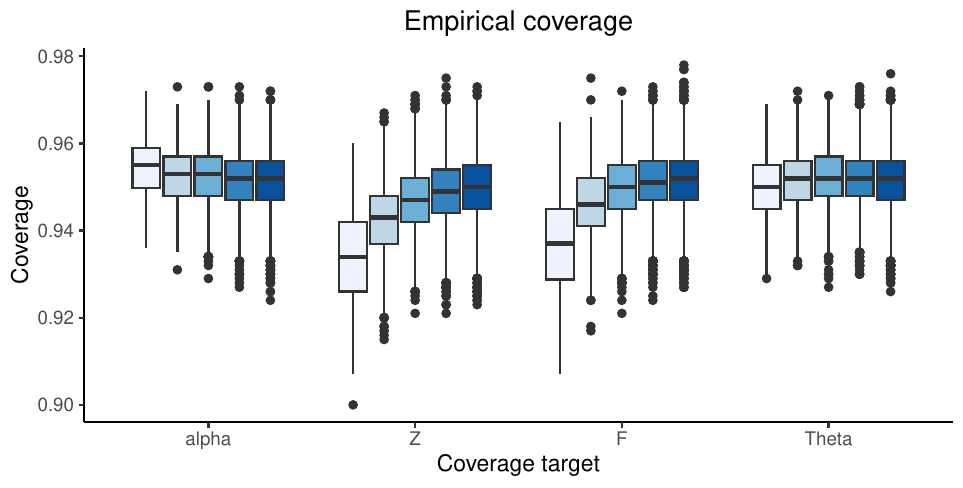} &
      \includegraphics[width=7.2cm]{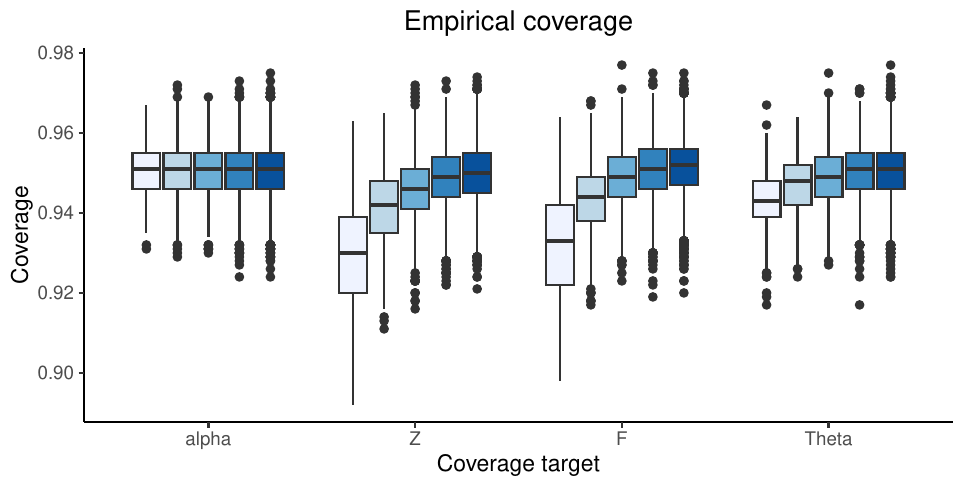} \\
      ~ & ~ \\ \vspace*{0.3cm}
      \includegraphics[width=7.2cm]{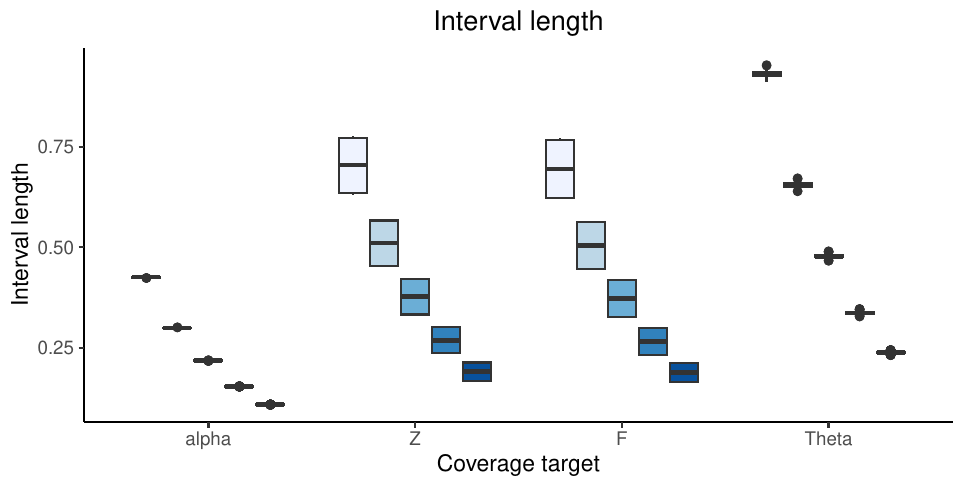} &
      \includegraphics[width=7.2cm]{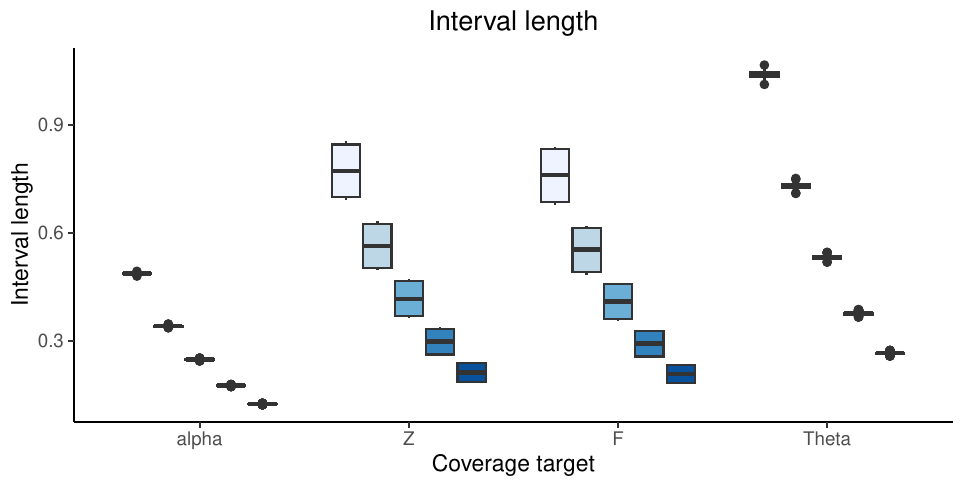} \\
      ~ & ~ \\
      (a) $\beta^*_{m,n} = 0$ &      (b) $\beta^*_{m,n} = -1$ 
      \end{tabular}
      \caption{\footnotesize{Empirical coverage for $\{\alpha^\dagger_i\}_{i\in[n]}$, $\{Z_{ik}^*\}_{i\in[n],k\in[2]}$,$ \{F_{jk}^*\}_{j\in[m],k\in[2]}$ and $\{\theta_{ii}^*\}_{i\in[m]}$, along with interval lengths. Each value comes from $1,000$ Monte Carlo repetitions. }}
      \label{fig:cov_all}
\end{figure}

\begin{figure}[t]
    \centering
      \setlength\tabcolsep{0pt}
	\renewcommand{\arraystretch}{-5}      
      \begin{tabular}{cc}   
      \multicolumn{2}{c}{\includegraphics[width = 8cm]{figs_sagfm/legend_inf.png}  } \vspace*{0.3cm} \\
      ~ & ~ \\ \vspace*{0.5cm}
      \includegraphics[width=7.5cm]{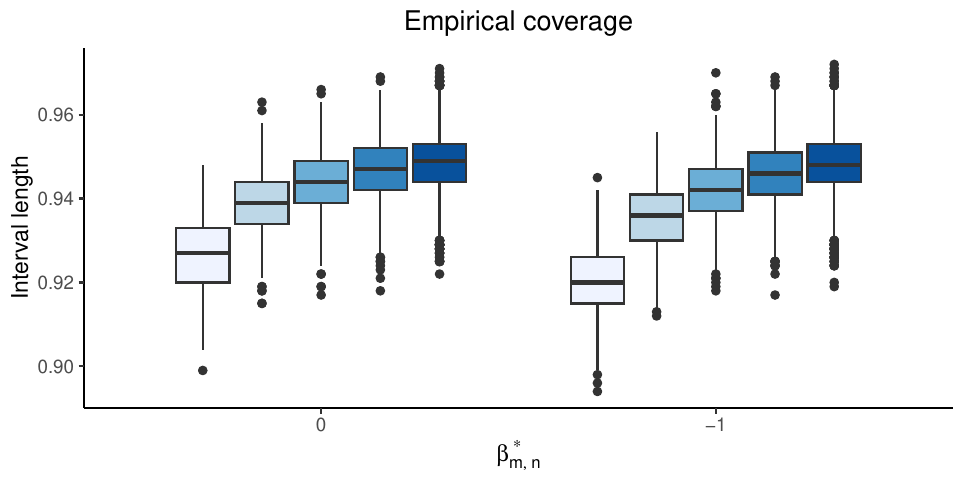} &
      \includegraphics[width=7.5cm]{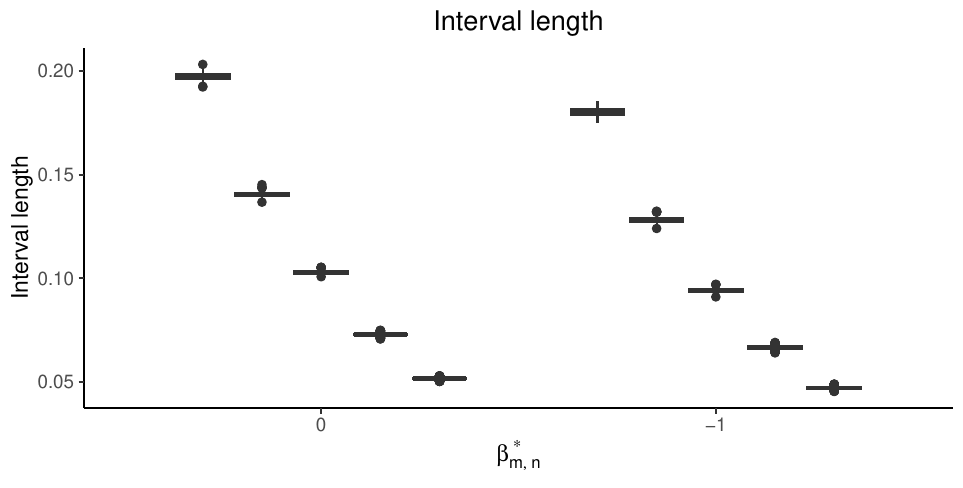} 
      \end{tabular}
      \caption{\footnotesize{Empirical coverage and interval length for $\{p_{ii}^*\}_{i\in[m]}$. Each value comes from $1,000$ Monte Carlo repetitions. }}
      \label{fig:cov_p}
\end{figure}

In this section, we evaluate the finite sample performance of the constructed confidence intervals for our model parameters, including the embedding vectors, the un-centered degree heterogeneity parameters,  $\theta_{ji}$'s, and  $p_{ji}$'s. We set $K=2$ and vary $m=n\in\{400,800,1500,3000,6000 \}$, $\beta^*_{m,n}\in\{0,-1\}$.  After generating $(\beta^*_{m,n}, \balpha^*, \bZ^*, \bF^*)$, $({\balpha}^{\dagger}, \bZ^*,\bF^*)$ are reconstructed accordingly. 
The $95\%$ confidence intervals are constructed following the methodologies outlined in Section \ref{sec:pmle}. 
Figure \ref{fig:cov_all} displays the empirical coverage values  and confidence interval lengths for  $(\balpha^\dagger,\bff^*, \bz^*, \bTheta^*)$ and Figure \ref{fig:cov_p} presents the results for $\bP^*$, with both figures considering a range of $(m,n)$ values. 
They are presented in two separate plots due to the differing scales of $(\balpha^\dagger,\bff^*, \bz^*, \bTheta^*)$ and $\bP^*$, making it more informative to display their interval lengths separately. In Figure \ref{fig:cov_all}, the upper box plots summarize empirical coverage values for each specified coverage target. For example, the ``F" coverage target at $n = 3000$ aggregates $6000$ empirical coverage values for $\{F_{jk}^*\}_{j\in[3000],k\in[2]}$. For the case where the coverage target is ``Theta" or the probability term, the box plots only display empirical coverage values for $\theta_{ji}^*$ with $j=i$, to avoid overburdening the plot software with all $(j,i)$ pairs. The empirical coverage values are based on 1000 Monte Carlo simulations. The box plots of the confidence interval lengths are constructed similarly.

We can see that {under both $\beta^*_{m,n} = 0$ and $\beta^*_{m,n} = -1$, the empirical coverage values for different coverage targets all approach the nominal $95\%$ level as $m$ and $n$ increase. In both cases, the empirical coverage of $\alpha_i^\dagger$ achieves $95\%$ with relatively small sample sizes, while the empirical coverage of the latent embeddings is initially below $95\%$ and reaches $95\%$ as the sample size increases, highlighting the difficulty of making inference in the latent space.
Furthermore, as $(m,n)$ increase, the lengths of the confidence intervals shrink notably. 
For fixed $\beta^*_{m,n}$, the interval length appears proportional to $(m \wedge n)^{-1/2}$. 
Moreover, the sparser the hypergraph, i.e., with smaller $\beta^*_{m,n}$, the wider the confidence intervals. 
These empirical findings are consistent with our theory.}

\section{The co-citation hypergraph}\label{sec:author}
 
In this section, we apply our latent embedding approach to a co-citation hypergraph. We use the Multi-Attribute Dataset on Statisticians (MADStat) collected and cleaned by \cite{ji2022co}. MADStat consists of citation and authorship information of 83,331 published papers with 47,311 authors, spanning from 1975 to 2015 and covering 36 statistical journals. We focus on papers published in the 36 journals from 1991 to 2000
and consider the top $3,000$ authors by  cited frequency within these years, which leaves us $14,984$ papers that cited these authors.
In the co-citation hypergraph, each author is a node, and each paper forms a hyperlink: if an author is cited by a paper, they appear on the corresponding hyperlink. 
That is, for the $j$th hyperlink (paper) $\by_j\in\{0,1\}^{3000}$, $y_{ji}=1$ if the $i$th author is cited by the $j$th paper and $y_{ji}=0$ otherwise.  

\begin{figure}[t]
    \centering
    \includegraphics[width = 0.9\textwidth]{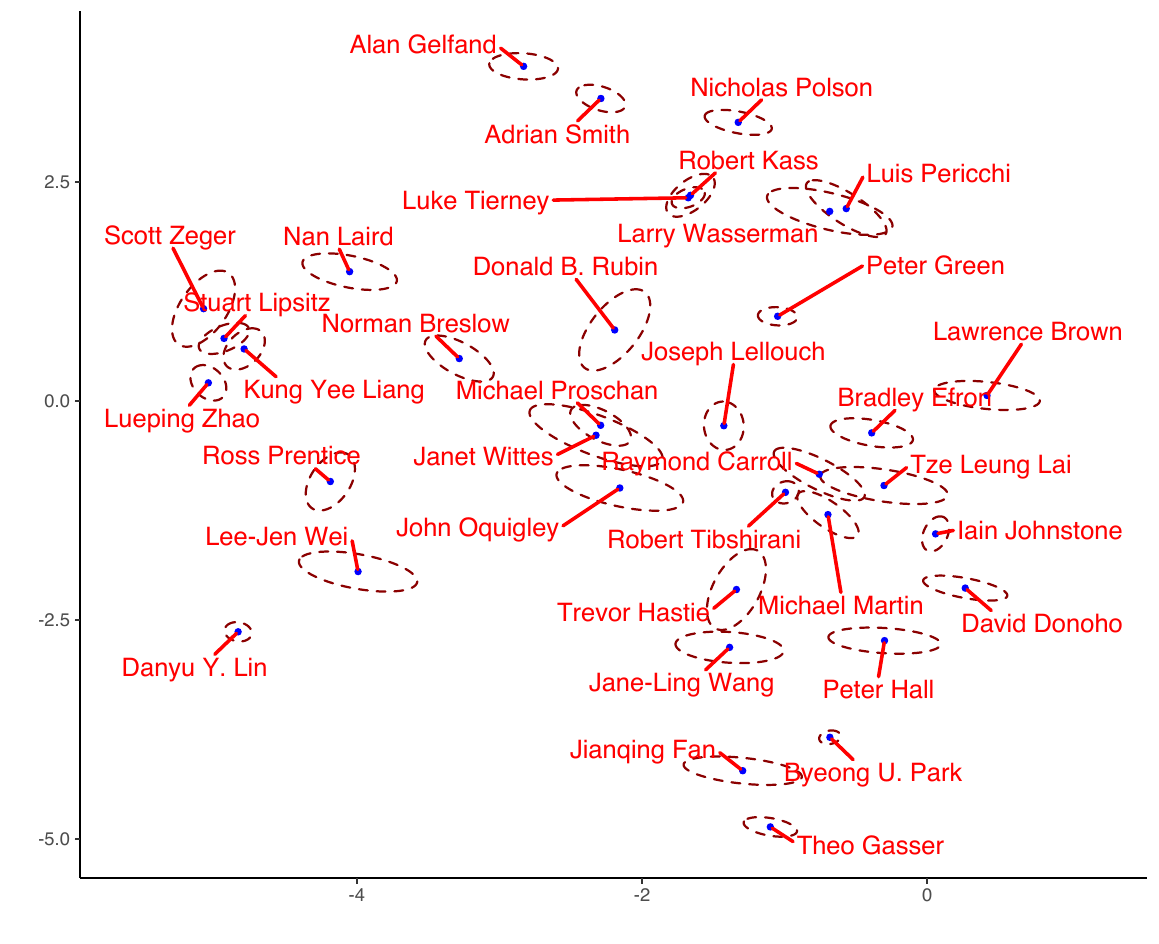}
    \caption{The author embedding vectors with $95\%$ confidence regions based on the co-citation hypergraph from 1991 to 2000. }
    \label{fig:regions_final}
\end{figure}

To obtain embedding vectors for the authors, we apply Algorithms 1 and 2 in the supplemental material to the $14,984$ hyperlinks of the $3,000$ authors, with the projection step based on Section \ref{sec:pmle}.
In Figure \ref{fig:regions_final}, we plot the embedding vectors along with $95\%$ confidence regions for a group of selected authors. The selected authors come from a joint set of authors with the highest estimated degree parameter values and a group of authors highlighted in the research map in \cite{ji2022co}.  The positioning of the embedding vectors shows relationship of recognized research interests, as evidenced by co-citation, among the authors between 1991 and 2000. For instance, the embedding vectors of Luke Tierney and Robert Kass are positioned almost identically, which indicates their possibly shared research interests from co-citation. Specifically, among the $14,984$ hyperlinks (papers), Tierney was cited in $349$ papers, while Kass received citations from $174$ papers, with $106$ of these papers citing both. Similarly, Larry Wasserman and Luis Pericchi have closely located embeddings with overlapping confidence regions. Wasserman was cited by 121 papers, Pericchi by 69, and they share 24 co-citations. 


Further, the sizes and orientations of the confidence regions could provide more insights on the hypergraph. From Theorem \ref{thm:limit_dist} and Corollary \ref{coro:var_limit}, the estimated covariance for author $i$'s embedding is $\big\{\big(\sum_{j=1}^m \sigma'(\hat{\theta}_{ji})\hat{\bq}_j\hat{\bq}_j^{\top} \big)^{-1}\big\}_{2:3,2:3}$, where $\hat{\bq}_j = (1, \hat{\bff}_j^{\top})^{\top}$ for $j\in[m]$ and $\{\bA\}_{2:3,2:3}$ denotes the sub-matrix of $\bA$ with columns and rows indexed in $\{2,3\}$. First consider a simplified case with $\hat{\theta}_{ji} = \theta_i$ for all $j\in[m]$; then we have 
\[
\big(\sum_{j=1}^m \sigma'(\hat{\theta}_{ji})\hat{\bq}_j\hat{\bq}_j^{\top} \big)^{-1} = \big\{\sigma'(\theta_{i}) \big\}^{-1} \begin{bmatrix}
    m^{-1} & \mathbf{0}_2^{\top} \\ \mathbf{0}_2 & \big(\sum_{j=1}^m\hat{\bff}_j\hat{\bff}_j^{\top}\big)^{-1}
\end{bmatrix}
\]
due to the fact that $\sum_{j=1}^m\hat{\bff}_j = \mathbf{0}_2$. Consequently, the estimated covariance for author $i$'s embedding would be $\bSigma_i:=\big\{\sigma'(\theta_{i})\big\}^{-1} \big(\sum_{j=1}^m\hat{\bff}_j\hat{\bff}_j^{\top}\big)^{-1}$. The magnitude of $\bSigma_i$ is determined by $m^{-1}$ times the inverse of the magnitude of the ``sample covariance'' quantity $m^{-1}\sum_{j=1}^m\hat{\bff}_j\hat{\bff}_j^{\top}$ and the magnitude of $\{\sigma'(\theta_i)\}^{-1}$. The orientation of $\bSigma_i$ is  the orientation of the ellipse of the   ``sample covariance'' $m^{-1}\sum_{j=1}^m\hat{\bff}_j\hat{\bff}_j^{\top}$, with major/minor axes being flipped. This is due to the fact that $m^{-1}\sum_{j=1}^m\hat{\bff}_j\hat{\bff}_j^{\top}$ shares the same eigen-space as $(m^{-1}\sum_{j=1}^m\hat{\bff}_j\hat{\bff}_j^{\top})^{-1}$ while the eigen-values are inversed.

\begin{figure}[ht]
      \centering
      \setlength\tabcolsep{0pt}
	\renewcommand{\arraystretch}{-5}      
      \begin{tabular}{ll}
       \includegraphics[width = 4.5cm]{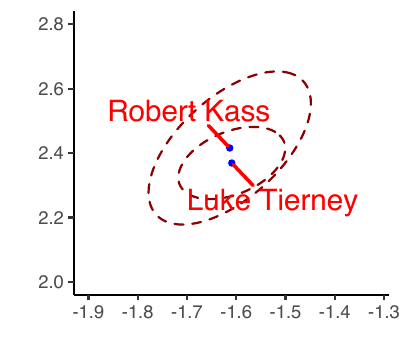} & \includegraphics[width = 4.5cm]{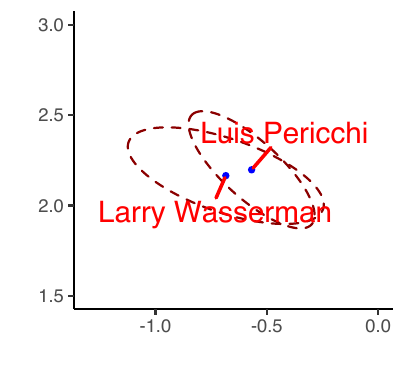} \\
      \includegraphics[width=7.425cm]{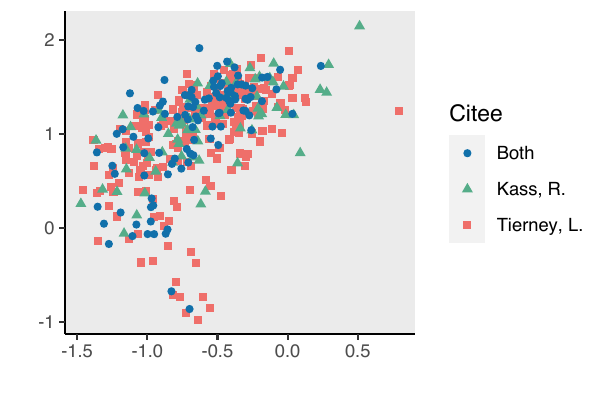} & \includegraphics[width=8.325cm]{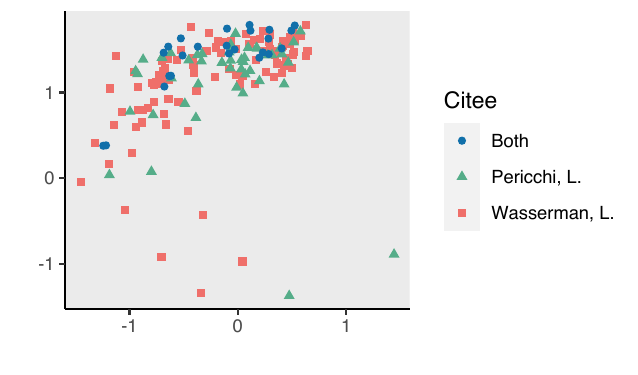}  
      \end{tabular}
      \caption{\footnotesize{Zoom-in versions of the embeddings in Figure \ref{fig:regions_final} for the four authors discussed and corresponding paper embeddings.}}
      \label{fig:paper_embedding}
\end{figure}

In practical scenarios where $\hat{\theta}_{ji}$ vary across different $j$, the estimated covariance matrices are the inverses of weighted second moments of $\hat{\bq}_j$'s, the analytic forms of which are rather challenging to evaluate directly.  We use Figure \ref{fig:paper_embedding} to explore the relationship between the shape of the confidence regions and the paper embeddings.   At the top of  Figure \ref{fig:paper_embedding}, we zoom in two parts in Figure \ref{fig:regions_final}. Each of these two parts highlights two author embeddings along with their confidence regions in the plots while omitting other authors. Beneath each zoomed-in part, we display the paper embeddings corresponding to the two authors. We then compare the sizes and orientations of the confidence regions for two pairs of authors: Tierney and Kass, and Pericchi and Wasserman, respectively.  Despite the similarity in the magnitude and distribution of paper embeddings for Tierney and Kass, there is a notable difference in the size of their confidence regions.  This is due to their different scales of $\sigma'(\hat{\theta}_{ji})$'s, which accord with different numbers of citations they received as discussed before. The median value of $\sigma'(\hat{\theta}_{ji})$ for Tierney is  $1.57\times 10^{-3}$, while the median value for Kass is $6.90\times 10^{-4}$. Such a difference in the scale of $\sigma'(\hat{\theta}_{ji})$'s leads to a larger confidence region for Kass's embedding.  Wasserman and Pericchi have confidence regions of similar sizes, while their orientations are different. This can be explained by their corresponding paper embeddings. The majority of their paper embeddings are closely located and similarly  distributed. However, a few paper embeddings at the bottom of the plot (indicated by red rectangulars) drift the general distribution of the embeddings of papers citing Wasserman counterclockwise, resulting in a slight rotation in his author embedding's confidence region. These observations shed some light on more understanding of the interactions among the vertices (authors) and their connections through individual hyperlinks.

\begin{figure}
    \centering
    \begin{tabular}{cc}
      \includegraphics[width = 5cm]{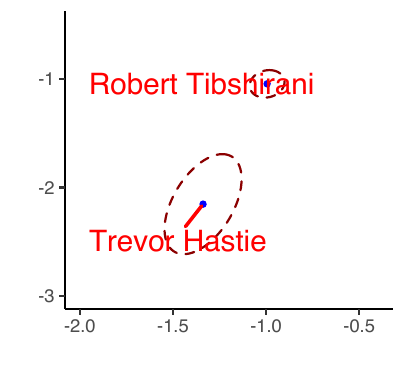}   &  \includegraphics[width = 7.5cm]{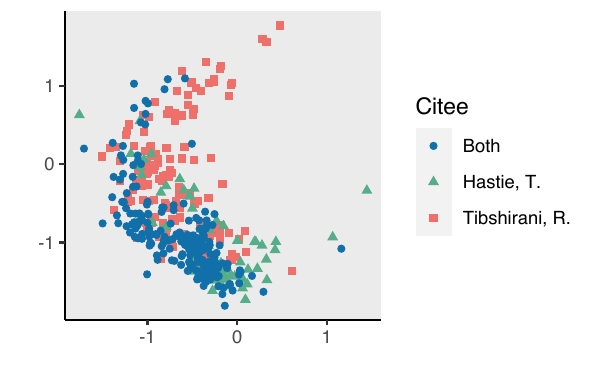} 
    \end{tabular}
    \caption{Author embeddings of Hastie T. and Thishirabi, R. and embeddings of papers citing them.}
    \label{fig:ht}
\end{figure}

In addition, another interesting observation is that the embeddings of Trevor Hastie and Robert Tibshirani in Figure \ref{fig:regions_final}, despite their frequent collaborations,  do not locate as close as one might expect. We zoom-in their author embeddings along with corresponding confidence regions   and plot the embeddings of papers citing their work between 1991 and 2000 in Figure \ref{fig:ht}.   Notably, a subset of these papers, particularly those positioned in the upper right part of the plot (marked via red rectangles), only cite Tibshirani but not Hastie. These papers all reference ``Efron, B. and Tibshirani, R. J. (1986) Bootstrap methods for standard errors, confidence intervals and other measures of statistical accuracy (with discussion). \textit{Statistical Science} $\mathbf{1}$(1):54-75'' \citep{efron1986bootstrap}, which is extensively cited in the data set. 
Interestingly, in the data set, none of the papers co-authored by Hastie  have ``bootstrap'' in their titles, whereas this term  appears multiple times in Tibshirani's co-authored papers. These observations unveil a difference in terms of research interests between Hastie and Tibshirani according to their co-citation information between 1991-2000, helping explain the different positioning of Hastie's and Tibshirani's embeddings.
Also notably, Tibshirani's embedding has a smaller confidence region compared to Hastie's. This is again due to different scales of their $\sigma'(\hat{\theta}_{ji})$'s. Specifically,  the median value
of $\sigma'(\hat{\theta}_{ji})$ for Hastie is $2.26\times 10^{-3}$, in contrast to Tibshirani's which is  
$1.09\times 10^{-2}$.   The different scales of ${\sigma}'(\hat{\theta}_{ji})$'s accord with their different frequencies of getting cited in the data set: Hastie by $283$ papers and Tibshirani by $347$.

\section{Discussion}\label{sec:discussion}

This paper introduces  a novel latent embedding approach for modeling high-dimensional general hypergraphs. The core objective of this methodology is to address pervasive and intricate challenges in hypergraph modeling: non-uniformity and multiplicity of  hyperlinks. While we consider embedding the network vertices and the hyperlinks in the Euclidean space in this paper, the latent embedding approach can serve as a general framework for modeling hypergraphs, with various choices of latent spaces, similarity measures, and activation functions. Under the Euclidean inner product considered in this paper, the embedding vectors from our methodology provide meaningful interpretations. Specifically,
the likelihood of any particular group of vertices appearing on a certain hyperlink is determined by the similarity between the hyperlink's embedding vector and the centroid of the vertices' embedding vectors.
To handle the complex nature of real-world hypergraphs, we further integrate degree heterogeneity parameters, which serve to adjust the vertices' different popularity, 
and include an order-adjusting parameter to model the overall order of the hyperlinks. The interpretability of our approach builds upon the integration of these parameters. 
Identifiability, estimation, and asymptotic behaviors are then studied within this framework. 

 In our theoretical analysis, we utilize the constraint sets in the estimation procedure and study a corresponding penalized maximum likelihood estimator (PMLE).  
 Demonstrating the equivalence between the constrained maximum likelihood estimator (CMLE) and the PMLE, the construction of the PMLE facilitates a more accurate characterization of the CMLE's properties, which yields the results in entry-wise consistency and asymptotic normality in this paper. In general, this Lagrangian multiplier type strategy has the potential for broad applications in a variety of constrained statistical estimation and inference problems. Moreover, our theoretical analysis studies the impact of the order-adjusting parameter on both estimation and inference procedures. We show that the exponential of the order-adjusting parameter is proportional to the expected order of the hyperlinks divided by the number of vertices. 
We require that the expected order of hyperlinks should be at least proportional to $n\cdot\log(m\vee n)/(m\wedge n)$ for $F$-consistent estimation and to  $n(m\vee n)^{\epsilon^*}/(m\wedge n)$ for some $\epsilon^*>0$ for entry-wise consistent estimation. For valid inference, we further need the expected order to be at least proportional to $ n(m+n)^{1+\epsilon^*}/(m\wedge n)^2$. These technical results are not only of theoretical interest, but also useful to guide practitioners on how and when to apply the latent embedding approach.  By understanding the conditions necessary for the effective application of the latent embedding approach, users are better equipped to leverage this methodology in appropriate contexts.

Building on the inference results of this work, various downstream tasks in hypergraph modeling, such as hyperlink prediction and community detection, can be handled. 
We leave detailed explorations on these downstream tasks for future research.  
Another important future work concerns variations of the latent embedding approach beyond the linear inner product framework. 
For instance, \cite{ma2020universal} studied a general model class adhering to the Schoenberg condition \citep{schoenberg1937certain,schoenberg1938metric}, which goes beyond the linear inner product-based latent space model for dyadic networks. While the utilization of linear inner product in \eqref{eq:hyper0} provides a clear and intuitive interpretation,  there is significant interest in developing and  assessing embedding methods using alternative inner product operations. Comparison between these operations with the linear inner product could bring new insights. Further appealing research avenues include investigating the choice of the link (activation) function and determining the optimal latent space dimension $K$. The selection of appropriate modeling approaches is also of great interest, for instance, adopting the study in
\cite{li2020network}.

\spacingset{1}
	\bibliographystyle{agsm}
	\bibliography{ref_jasa}

@article{wang2022maximum,
  title={Maximum likelihood estimation and inference for high dimensional generalized factor models with application to factor-augmented regressions},
  author={Wang, Fa},
  journal={Journal of Econometrics},
  volume={229},
  number={1},
  pages={180--200},
  year={2022},
  publisher={Elsevier}
}

@article{ke2019community,
  title={Community detection for hypergraph networks via regularized tensor power iteration},
  author={Ke, Zheng Tracy and Shi, Feng and Xia, Dong},
  journal={arXiv preprint arXiv:1909.06503},
  year={2019}
}

@article{zhen2022community,
  title={Community detection in general hypergraph via graph embedding},
  author={Zhen, Yaoming and Wang, Junhui},
  journal={Journal of the American Statistical Association},
  volume={118},
  pages={1620-1629},
  year={2022},
  publisher={Taylor \& Francis}
}

@article{hoff2002latent,
  title={Latent space approaches to social network analysis},
  author={Hoff, Peter D and Raftery, Adrian E and Handcock, Mark S},
  journal={Journal of the American Statistical Association},
  volume={97},
  number={460},
  pages={1090--1098},
  year={2002},
  publisher={Taylor \& Francis}
}

@article{kumar2018hypergraph,
  title={Hypergraph clustering: a modularity maximization approach},
  author={Kumar, Tarun and Vaidyanathan, Sankaran and Ananthapadmanabhan, Harini and Parthasarathy, Srinivasan and Ravindran, Balaraman},
  journal={arXiv preprint arXiv:1812.10869},
  year={2018}
}

@article{lee2020robust,
  title={Robust hypergraph clustering via convex relaxation of truncated MLE},
  author={Lee, Jeonghwan and Kim, Daesung and Chung, Hye Won},
  journal={IEEE Journal on Selected Areas in Information Theory},
  volume={1},
  number={3},
  pages={613--631},
  year={2020},
  publisher={IEEE}
}

@article{ghoshdastidar2017consistency,
  title={Consistency of spectral hypergraph partitioning under planted partition model},
  author={Ghoshdastidar, Debarghya and Dukkipati, Ambedkar},
  journal={The Annals of Statistics},
  volume={45},
  pages={289-315},
  year={2017}
}

@article{ji2016coauthorship,
  title={Coauthorship and citation networks for statisticians},
  author={Ji, Pengsheng and Jin, Jiashun},
  journal={The Annals of Applied Statistics},
  volume ={10},
  pages={1779-1812},
  year={2016}
}

@article{ji2022co,
  title={Co-citation and co-authorship networks of statisticians},
  author={Ji, Pengsheng and Jin, Jiashun and Ke, Zheng Tracy and Li, Wanshan},
  journal={Journal of Business \& Economic Statistics},
  volume={40},
  number={2},
  pages={469--485},
  year={2022},
  publisher={Taylor \& Francis}
}

@article{newman2001structure,
  title={The structure of scientific collaboration networks},
  author={Newman, Mark EJ},
  journal={Proceedings of the National Academy of Sciences},
  volume={98},
  number={2},
  pages={404--409},
  year={2001},
  publisher={National Acad Sciences}
}

@article{fowler2006connecting,
  title={Connecting the congress: a study of cosponsorship networks},
  author={Fowler, James H},
  journal={Political Analysis},
  volume={14},
  number={4},
  pages={456--487},
  year={2006},
  publisher={Cambridge University Press}
}

@article{lee2017time,
  title={Time-dependent community structure in legislation cosponsorship networks in the Congress of the Republic of Peru},
  author={Lee, Sang Hoon and Magallanes, Jos{\'e} Manuel and Porter, Mason A},
  journal={Journal of Complex Networks},
  volume={5},
  number={1},
  pages={127--144},
  year={2017},
  publisher={OUP}
}

@inproceedings{ghoshdastidar2015provable,
  title={A provable generalized tensor spectral method for uniform hypergraph partitioning},
  author={Ghoshdastidar, Debarghya and Dukkipati, Ambedkar},
  booktitle={International Conference on Machine Learning},
  pages={400--409},
  year={2015},
  organization={PMLR}
}

@article{razick2008irefindex,
  title={iRefIndex: a consolidated protein interaction database with provenance},
  author={Razick, Sabry and Magklaras, George and Donaldson, Ian M},
  journal={BMC Bioinformatics},
  volume={9},
  number={1},
  pages={1--19},
  year={2008},
  publisher={BioMed Central}
}

@article{yuan2021high,
  title={High-order joint embedding for multi-level link prediction},
  author={Yuan, Yubai and Qu, Annie},
  journal={Journal of the American Statistical Association},
  volume={118},
  pages={1692-1706},
  year={2021},
  publisher={Taylor \& Francis}
}

@article{ma2020universal,
  title={Universal latent space model fitting for large networks with edge covariates},
  author={Ma, Zhuang and Ma, Zongming and Yuan, Hongsong},
  journal={The Journal of Machine Learning Research},
  volume={21},
  number={1},
  pages={86--152},
  year={2020},
  publisher={JMLRORG}
}

@article{chatterjee2015matrix,
  title={Matrix estimation by universal singular value thresholding},
  author={Chatterjee, Sourav},
  journal={The Annals of Statistics}, 
  volume = {43},
  pages = {177-214},
  year={2015}
}

@article{schoenberg1937certain,
  title={On certain metric spaces arising from Euclidean spaces by a change of metric and their imbedding in Hilbert space},
  author={Schoenberg, Isaac J},
  journal={Annals of Mathematics},
  volume = {38},
  pages={787--793},
  year={1937},
  publisher={JSTOR}
}

@article{schoenberg1938metric,
  title={Metric spaces and positive definite functions},
  author={Schoenberg, Isaac J},
  journal={Transactions of the American Mathematical Society},
  volume={44},
  number={3},
  pages={522--536},
  year={1938}
}

@article{silvey1959lagrangian,
  title={The Lagrangian multiplier test},
  author={Silvey, Samuel D},
  journal={The Annals of Mathematical Statistics},
  volume={30},
  number={2},
  pages={389--407},
  year={1959},
  publisher={JSTOR}
}

@article{el1994wald,
  title={On the {W}ald, {L}agrangian multiplier and likelihood ratio tests when the information matrix is singular},
  author={El-Helbawy, Abdalla T and Hassan, Tawfik},
  journal={Journal of the Italian Statistical Society},
  volume={3},
  pages={51--60},
  year={1994},
  publisher={Springer}
}

@article{yuan2022testing,
  title={Testing community structure for hypergraphs},
  author={Yuan, Mingao and Liu, Ruiqi and Feng, Yang and Shang, Zuofeng},
  journal={The Annals of Statistics},
  volume={50},
  number={1},
  pages={147--169},
  year={2022},
  publisher={Institute of Mathematical Statistics}
}

@article{li2020network,
  title={Network cross-validation by edge sampling},
  author={Li, Tianxi and Levina, Elizaveta and Zhu, Ji},
  journal={Biometrika},
  volume={107},
  number={2},
  pages={257--276},
  year={2020},
  publisher={Oxford University Press}
}

@book{newman2018networks,
  title={Networks},
  author={Newman, Mark},
  year={2018},
  publisher={Oxford university press}
}

@article{holland1983stochastic,
  title={Stochastic blockmodels: First steps},
  author={Holland, Paul W and Laskey, Kathryn Blackmond and Leinhardt, Samuel},
  journal={Social networks},
  volume={5},
  number={2},
  pages={109--137},
  year={1983},
  publisher={Elsevier}
}

@article{efron1986bootstrap,
  title={Bootstrap Methods for Standard Errors, Confidence Intervals, and Other Measures of Statistical Accuracy},
  author={Efron, B and Tibshirani, R},
  journal={Statistical Science},
  volume={1},
  number={1},
  pages={54--75},
  year={1986}
}

@article{mukherjee2018detection,
  title={Detection thresholds for the $\beta$-model on sparse graphs},
  author={Mukherjee, Rajarshi and Mukherjee, Sumit and Sen, Subhabrata},
  journal={The Annals of Statistics},
  volume={46},
  number={3},
  pages={1288--1317},
  year={2018},
  publisher={JSTOR}
}

@article{turnbull2024latent,
  title={Latent space modeling of hypergraph data},
  author={Turnbull, Kathryn and Lunag{\'o}mez, Sim{\'o}n and Nemeth, Christopher and Airoldi, Edoardo},
  journal={Journal of the American Statistical Association},
  volume={119},
  number={548},
  pages={2634--2646},
  year={2024},
  publisher={Taylor \& Francis}
}

@article{nandy2024degree,
  title={Degree heterogeneity in higher-order networks: Inference in the hypergraph $\beta$-model},
  author={Nandy, Sagnik and Bhattacharya, Bhaswar B},
  journal={IEEE Transactions on Information Theory},
  year={2024},
  publisher={IEEE}
}

@inproceedings{stasi2014beta,
  title={$\beta$ models for random hypergraphs with a given degree sequence},
  author={Stasi, Despina and Sadeghi, Kayvan and Rinaldo, Alessandro and Petrovic, Sonja and Fienberg, Stephen},
  booktitle={COMPSTAT 2014 21st International Conference on Computational Statistics},
  pages={593},
  year={2014},
  organization={Citeseer}
}

@article{chatterjee2011random,
  title={Random graphs with a given degree sequence},
  author={Chatterjee, Sourav and Diaconis, Persi and Sly, Allan},
  journal={The Annals of Applied Probability},
  volume={21},
  number={4},
  pages={1400--1435},
  year={2011},
  publisher={Citeseer}
}

\end{document}